\documentclass[a4paper,showpacs,twocolumn,prd,aps,10pt,amsmath,amssymb]{revtex4-1}
\pdfoutput=1
\usepackage{amsmath,amsthm}
\usepackage{array}
\usepackage{mathtools}
\usepackage{bm}
\usepackage{graphicx}
\usepackage{tensor}
\usepackage[hyperindex,breaklinks]{hyperref}
\usepackage{tikz}
\usetikzlibrary{arrows,positioning,shapes}

\allowdisplaybreaks[1] 

\newcommand{\del}{\partial}
\newcommand{\zcpm}{z_{c\pm}}
\newcommand{\bpm}{B_{c\pm}}
\newcommand{\cpm}{C_{c\pm}}
\newcommand{\bdu}{b \cdot u}
\newcommand{\bdh}{b \cdot \hat{u}}
\newcommand{\Li}{\operatorname{Li}}
\newcommand{\LL}{\operatorname{L}}
\newcommand{\zs}{\zeta_s}
\newcommand{\zt}{\zeta_t}
\newcommand{\sgn}{\text{sgn}}
\newcommand{\vk}{\mathbf{k}}
\newcommand{\vq}{\mathbf{q}}
\newcommand{\vp}{\mathbf{p}}
\newcommand{\vx}{\mathbf{x}}
\newcommand{\tr}{\operatorname{tr}}
\newcommand{\T}{\mathsf{T}}
\newcommand{\oo}{\infty}
\newcommand{\x}{\hat{x}}
\newcommand{\expr}{\langle \tilde{r}^2 \rangle}

\renewcommand{\O}{\mathcal{O}}
\renewcommand{\d}{\mathrm{d}}
\renewcommand{\u}{{\hat{u}}}
\renewcommand{\Re}{\operatorname{Re}}

\begin{document}

\title{Quantum astrometric observables. II. Time delay in linearized quantum gravity}

\author{B\'eatrice Bonga}
\altaffiliation[Presently at ]{[Penn State]}
\email{bpb165@psu.edu}
\author{Igor Khavkine}
\email{i.khavkine@uu.nl}
\affiliation{Institute for Theoretical Physics, Utrecht University\\
Leuvenlaan 4, NL--3584 CE Utrecht, The Netherlands}
\date{\today}

\begin{abstract}
A clock synchronization thought experiment is modeled by a
diffeomorphism invariant ``time delay'' observable. In a sense, this
observable probes the causal structure of the ambient Lorentzian
spacetime. Thus, upon quantization, it is sensitive to the long expected
smearing of the light cone by vacuum fluctuations in quantum gravity.
After perturbative linearization, its mean and variance are computed in
the Minkowski Fock vacuum of linearized gravity. The na\"ive divergence of
the variance is meaningfully regularized by a length scale $\mu$, the
physical detector resolution. This is the first time vacuum fluctuations
have been fully taken into account in a similar calculation. Despite
some drawbacks this calculation provides a useful template for the study
of a large class of similar observables in quantum gravity. Due to their
large volume, intermediate calculations were performed using computer
algebra software. The resulting variance scales like $(s \ell_p/\mu)^2$,
where $\ell_p$ is the Planck length and $s$ is the distance scale
separating the (``lab'' and ``probe'') clocks. Additionally, the
variance depends on the relative velocity of the lab and the probe,
diverging for low velocities. This puzzling behavior may be due to an
oversimplified detector resolution model or a neglected second-order
term in the time delay.
\end{abstract}

\maketitle

\section{Introduction}
In a previous paper~\cite{khavkine}, one of us proposed a gauge
invariant and operationally meaningful observable, the \emph{time
delay}, as a test case for practical calculations in perturbative
quantum gravity and as a probe of the causal structure of both classical
and quantum gravity. As is well known, the issue of gauge invariant
(diffeomorphism invariant) observables is central in the physical
interpretation of relativistic gravity as well as in its
quantization~\cite{bergmann-obsv,rovelli-obsv,lp-obsv,dittrich-obsv,dittrich-tambo,pss-obsv}.
The definition of the \emph{time delay} is inspired by classical
relativistic astrometry~\cite{soffel-astro}.  Thus, in the quantum
context, it can be thought of as a member of a larger class of so-called
\emph{quantum astrometric observables}.

A detailed discussion of our approach to the question of observables in
both classical and quantum gravity can be found in~\cite{khavkine}. There, the
time delay was defined using an implicit operational description and
explicitly computed at linear perturbative order. Two exact inequalities
were also proven, demonstrating that the causal structure of a
Lorentzian metric imposes strict bounds on its values. Finally, a sketch
of a calculation of the variance of the time delay in the Minkowski
linearized quantum gravitational vacuum was given. The sketch pointed
out that the additional physical input of a finite measurement
resolution was necessary to obtain a finite result. However, the details
of the calculation, besides a simple dimensional analysis estimate, were
deferred. This calculation is presented in detail in this work, which is
based on the MSc thesis of one of the authors~\cite{bb-thesis}.

The calculation is in some ways significantly different from standard
quantum field theory calculations, which accounts for its complexity,
because it uses explicitly nonlocal observables rather than those
locally defined from the field operators or their Fourier transforms. We
recall that some similar calculations by other authors can be found
in~\cite{woodard-thesis,tsamis-woodard,ford-lightcone,ford-top,ford-focus,ford-angle,ohlmeyer,roura}.
The calculation in~\cite{woodard-thesis,tsamis-woodard} is in some ways
more complex and sophisticated, but the methods and focus of the result
are substantially different: they used an expansion to quadratic order,
dimensional regularization, and focused on the resulting regulated
divergences. The calculations
in~\cite{ford-lightcone,ford-top,ford-focus,ford-angle} have a greater
breadth in the choice of observables and vacua, but neglected important
issues: their results are somewhat difficult to disentangle from the
choice of gauge and the quantum fluctuations due to the Poincar\'e
invariant Fock vacuum proper were left uncomputed (as opposed to
additional thermal, squeezed or extra dimensional effects). The work
in~\cite{ohlmeyer} was technically similar, but focused on lengths of
spacelike segments and did not supply a plausible phenomenological
interpretation. The unpublished work of~\cite{roura} is most similar,
but makes significantly different technical choices and is restricted to
a limited choice of experimental geometries.

Our work is the first to compute the finite quantum variance
(regularized by a finite measurement resolution scale) of a quantum
astrometric observable in the Poincar\'e invariant Fock vacuum of
linearized quantum gravity; the observable is the time delay, which is
interesting because it is sensitive to the quantum fluctuations of the
light cones %
	\footnote{The recent work~\cite{leonard-woodard} also computes a fully
	renormalized observable sensitive to light cone fluctuations: the
	electromagnetic vacuum polarization. However, its phenomenological
	interpretation is less clear.}.
Moreover, several technical choices make it of wider interest. Since the
calculation is carried out entirely in position space, the qualitative
behavior of various singular integrals is expected to generalize to
calculations on a curved (background) spacetime. Also, (linearized)
gauge invariance of the calculation is manifest. Finally, the tools
constructed in its course, allow a straightforward generalization to
more complicated experimental geometries.

Unfortunately, some of the technical choices are not without drawbacks.
The choice of the family of detector resolution profiles explicitly
breaks Lorentz invariance (by treating the lab's reference frame as
preferred). Additionally, to be truly accurate to order $\ell_p^2$ (Planck length
squared), the linear order expression for the time delay that we used is
not sufficient and the quadratic order should also be included. Both of
these choices were made for the pragmatic reason of making the
complexity of the calculation manageable. Despite these limitations, we
believe this calculation can serve as a useful template for practical
calculations with quantum astrometric observables and can give
qualitative (though detailed) information about the expected results.

At this point, it should be emphasized that, in any realistic
experimental setup, there will be many sources of fluctuations,
including quantum fluctuations in the internal experimental apparatus
degrees of freedom. These fluctuations have been examined by many
authors~\cite{salecker-wigner,gambini-pullin}. Our calculations, on the
other hand, concentrate on the contribution to these fluctuations due
purely to quantum gravitational effects. Other fluctuation sources are
often found to have amplitudes exceeding Planck scales, while our
results show that amplitude of quantum gravitational fluctuations are,
as expected, set by the Planck scale. So, the quantum gravitational
fluctuations are rarely expected to constitute the primary signal.
However, they are worth examining for two reasons. First, it is not a
priori excluded that quantum gravitational fluctuations could constitute
a subleading but detectable contribution to the signal, especially if
some enhancement is possible that would remain unguessed unless the
actual calculation were performed. Second, it is worth understanding
these quantum gravitational fluctuations purely theoretically, as they
constitute a physical effect that is in principle different from those
in nongravitational systems, since they are produced in part by
quantum fluctuations in what we consider to be causal structure in
spacetime.

In Sec.~\ref{sectimedelay} we briefly recall from~\cite{khavkine} the definition
of the time delay observable and its main properties.
Section~\ref{secchoices} explicitly lists the technical choices determining
the result, together with the rationale behind them, and outlines the
strategy of the main calculation. The bulk of the computation is
performed with the aid of a computer, with the technical details of the
algorithm given in Sec.~\ref{seccalc}. For the actual computer code with
usage instructions see %
	\footnote{See Ancillary Files at \url{http://arxiv.org/src/1307.0256/anc} for the
	\textit{Mathematica} computer code reproducing the results of our
	Sec.~\ref{secresults}.}.
We present the results in Sec.~\ref{secresults} and conclude with a
discussion in Sec.~\ref{secdisc}. Appendices~\ref{apppertsol}
and~\ref{applinobsv} give details of the perturbative solution of the
geodesic equation.  Appendix~\ref{apptwopoint} justifies our form of the
graviton two-point function. And Appendix~\ref{apppartialcheck} shows some
manual calculations used for checking our computer code.

\section{The time delay observable}
\label{sectimedelay}
\subsection{Operational definition}
Here we briefly introduce the \emph{time delay} observable and summarize
its most relevant properties. A more extensive discussion of the problem
of observables in General Relativity and how the time delay fits into it
can be found in~\cite{khavkine}. 

\begin{figure}
\centering
\includegraphics[scale=1]{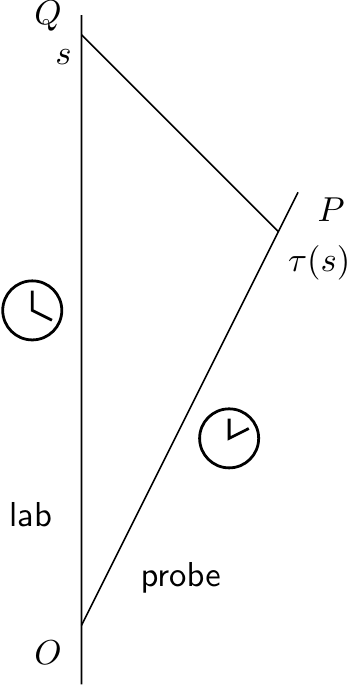} 
\caption{Geometry of the experimental protocol with the synchronization/ejection point $O$, the signal emission point $P$ at time $\tau(s)$ after emission and the signal reception point $Q$ at time $s$ after emission.}
\label{figexpsetup}
\end{figure}

We shall construct an observable by specifying a (thought) experiment
protocol (Fig.~\ref{figexpsetup}) and carefully constructing a
mathematical model of it. Since it is very difficult to imagine an
experiment executed by purely gravitational degrees of freedom, we must
introduce a minimal amount of matter content, just enough for an
idealized model of the experimental apparatus.

Consider a laboratory in inertial motion (free fall). The laboratory
carries a clock that measures the proper time along its trajectory. The
laboratory also carries an orthogonal frame, which is
parallel-transported along the lab's worldline. (The frame could be
Fermi-Walker transported if the motion were not inertial.) At a moment
of the experimenter's choosing, the lab ejects a probe in a
predetermined direction, fixed with respect to the lab's orthogonal
frame and with a predetermined relative velocity. The probe then
continues to move inertially and carries its own proper time clock. The
two clocks are synchronized to $0$ at the ejection event $O$.  After
ejection, the probe continuously broadcasts its own proper time
(time-stamped signals), in all directions using an electromagnetic signal. At a 
predetermined proper time interval $s$ after ejection, event $Q$, the lab 
records the probe signal and its emission time stamp $\tau(s)$, sent from 
event $P$.  Call $s$ the \emph{reception time}, $\tau(s)$ the 
\emph{emission time} and the difference,
\begin{equation}\label{eqemtime}
	\delta\tau(s)=s-\tau(s)
\end{equation}
the \emph{time delay}.

To model this protocol mathematically, we introduce the notion of a
\emph{lab-equipped spacetime} $(M,g,O,\hat{e}^a_i)$, which consists of
an oriented manifold $M$, with time oriented Lorentzian metric $g$, a
point $O\in M$ and an oriented orthonormal frame $\hat{e}^a_i\in T_O M$,
with $\hat{e}^a_0$ timelike and future oriented. The point $O$ is
identified with the \emph{probe ejection event}, while $\hat{e}_0^a$ is
tangent to the lab worldline. The probe worldline is tangent to a vector
$v^a = v^i \hat{e}_i^a$, whose components are specified with respect to
the tetrad at $O$ (the \emph{lab frame}). For a fixed relative probe
velocity $v^i$ and a fixed reception time $s$, once a lab-equipped
spacetime is given, it is a matter of solving the appropriate geodesic
equations to calculate the emission time $\tau_v(s)$ or the time delay
$\delta\tau_v(s)$. In the remaining, the explicit dependence of the time delay on $v$ and $s$ is omitted when the context is clear. Manifestly, both are invariant under diffeomorphisms
that simultaneously act on all components of the lab-equipped spacetime
data. It is worth noting that the time delay satisfies interesting
inequalities related to the causal structure of Lorentzian metrics. We
will not expand on this remark in this work, but refer the reader to
Secs.~IV and VII~B of~\cite{khavkine} for more details.

\subsection{Overview of the calculational procedure}
\label{subsecoverviewprocedure}
Unfortunately, the above definition, though exact and conceptually
clear, is not very useful in practical calculations. For that purpose,
we suppose that the spacetime $(M,g)$ is a small perturbation on top of
Minkowski space. We find an explicit linearized expression for the
time delay, as a linear function of the \emph{graviton field} (the
deviation of $g$ from the Minkowski metric). This linearized expression
will then be used to quantize the observable, by replacing the
classical graviton field with a smeared version of the quantum graviton
field (see Secs.~\ref{subsecquant} and~\ref{subsecsmearing}).

The Poincar\'e invariant Fock vacuum is chosen as the quantum
gravitational vacuum and the quantum time delay observable is evaluated
with respect to this vacuum.  Since the Fock vacuum is Gaussian with
respect to any observable that is linear in the graviton field, the
focus is on calculating the mean and the variance of our quantum
observable as this captures all the information about its quantum
measurements. The mean is the same as the classical Minkowski space
expression to linear order in the graviton field, however, the variance
$\langle0|\hat{\tau}(s)^2|0\rangle$ is more complicated and will be
calculated from the expectation value of the square of the quantized
linear correction to the time delay, which we denote by
$r[\hat{h}]$.

We derive an analytic expression for this quantum variance (see
Sec.\ref{subsecmaster}), which consists of 45 terms of which each
contains so-called smeared segment integrals that are composed of
one-dimensional integrals along the worldlines of the lab and the probe
and four-dimensional integrals over a smearing function.  This smearing
function smears the graviton field and guarantees that the quantum
variance is finite and can physically be interpreted as modeling the
detector sensitivity (see Sec.~\ref{subsecsmearing}).  We then take a
pragmatic view and chose to work not with a generic smearing, but with
one that extends only in the plane orthogonal to the lab worldline, with
spherical symmetry within it. At this point we resort to the use of
hybrid numerical-analytical calculations automated using computer
algebra software (\textsc{Mathematica}~8.0). The integration along the
geodesic segments and the angular smearing integrals are computed first
and then tabulated. After this, the remaining smearing is carried out to
obtain the quantum variance of the time delay for an arbitrary shape of
the triangle as described in the experimental protocol. The details of
the computer calculation are described in the rest of Sec.~\ref{seccalc}
and the results are reported in Sec.~\ref{secresults}.

\subsection{Linearized expression}
\label{subseclinearized}
To give the explicit linearized formula, we need some notation that is
introduced in Appendix~\ref{applinobsv}. Note that we use $h$ to denote
the graviton field and perform all index contractions using the
Minkowski metric. We parametrize the linear correction to the emission
time as
\begin{align}
	\tau(s) &= \tau_\mathrm{cl}(s) (1 + r[h]) + \O(h^2) , \\
\label{eqimplicitr}
	r[h] &= \sum_{KnX} r^{Kij}_{n X} \int^{(n)}_{X} \nabla_K h_{(ij)},
\end{align}
where $\int^{(n)}_X$ denotes an affinely $[0,1]$-parametrized,
$n$-iterated integral over a segment $X$. The summation is carried
out over the segments $X$, the integral iteration number $n$ and the
multi-indices $K$, with $r^{Kij}_{n X}$ some tensor coefficients to be
specified. An ordinary integral is zero-iterated $\int^{(0)}\d{t}\, f(t)
= \int_0^1 \d{t}\, f(t)$, while a one-iterated integral is $\int^{(1)}\d{t} f(t) =
\int_0^1\d{t}\int_0^t\d{t'}\, f(t')$. The multi-index $K=(i_1i_2\cdots
i_{|K|})$ defines the differential operator $\nabla_K = \del_{i_0}
\del_{i_1} \cdots \del_{i_{|K|}}$. The segments range over $X=U,V,W$,
which label the sides of the geodesic triangle defined in Minkowski
space by the time delay measurement protocol, illustrated in
Figs.~\ref{figexpsetup} and~\ref{triang-geom-H}. The vectors
corresponding to each segment are $U^a=-su^a$, $V^a=tv^a$ and $W^a=w^a$.
The vector $w^a$ is null, while $u^a$ and $v^a$ are future pointing,
timelike unit vectors, representing, respectively, the velocities of the
lab and probe worldlines. The probe ejection velocity is parametrized by
the rapidity $\theta$, which is defined by $u\cdot v = -\cosh\theta$.
The nonvanishing coefficient tensors $r^{Kij}_{n X}$ (namely, the
restriction to the ranges $n=0,1$ and $|K|=0,1$) can be read off
directly from the following explicit formula, which is obtained by
explicitly expanding the sums of the more structured
expression~\eqref{eqr-def}--\eqref{J-expr},
\begin{widetext}
\begin{align}
\notag
	r[h]
	&= \frac{1}{\tau_\mathrm{cl}(s) v\cdot w}
	\left(
		  2 W^{[i} U^{j]} V^k \int_V\d{t}\,\del_i h_{(kj)}
		+ 2 W^{[i} U^{j]} W^k \int_W\d{t}\,\del_i h_{(kj)}
		+ 2 W^{[i} U^{j]} U^k \int_U\d{t}\,\del_i h_{(kj)} \right. \\
\notag & \quad {}
		+ W^i V^j \int_V\d{t}\, h_{(ij)}
		- 2 W^{[i} V^{j]} V^k \int_V^{(1)}\d{t}\, \del_i h_{(kj)}
		+ W^i W^j \int_W\d{t}\, h_{(ij)}
		+ W^i U^j \int_U\d{t}\, h_{(ij)} \\
& \quad \left. {}
		- 2 W^{[i} U^{j]} U^k \int_U^{(1)}\d{t}\, \del_i h_{(kj)}
		- 2 W^{[i} U^{j]} W^k \int_W\d{t}\, \del_i h_{(kj)}
		- 2 W^{[i} U^{j]} V^k \int_V\d{t}\, \del_i h_{(kj)}
	\right) , \label{eqexplicitr}
\end{align}
\end{widetext}
where $\tau_\mathrm{cl}(s) = se^{-\theta}$ is the time delay computed in
Minkowski space, as in Eq.~\eqref{taucl}.

\subsection{Quantization}
\label{subsecquant}
The linearized gravitational field can be quantized fairly
straightforwardly, for instance, by using a complete gauge fixing and
constructing a Poincar\'e invariant Fock vacuum (see
Appendix~\ref{apptwopoint} for details). The quantization is completely
specified by the (Wightman) two-point function $\langle
\hat{h}(x)\hat{h}(y) \rangle$, where $\hat{h}(x)$ is the quantized field
corresponding to $h(x)$. In a standard way, using Wick's theorem, the
expectation value of any quantum observable can be expressed as a
function of $\langle \hat{h}(x)\hat{h}(y) \rangle$. We are ultimately
interested in computing the vacuum fluctuation in the quantized emission
time observable $\hat{\tau}(s)$, which in our approximation reduces to
computing the expectation value of the square of the quantized linear
correction $\widehat{r[h]}$. The latter quantity is expressible in terms
of the (Hadamard) two-point function
\begin{equation}\label{eqhadamarddef}
	\langle \{ \hat{h}(x) , \hat{h}(y) \} \rangle
	= \langle \hat{h}(x)\hat{h}(y) + \hat{h}(x)\hat{h}(y) \rangle
	\sim \frac{\ell_p^2}{(x-y)^2} ,
\end{equation}
whose precise form depends on the choice of gauge, but the
displayed singular term appears generically.

Since $r[h]$ is linear in the graviton field, the simplest quantization
prescription is to replace every occurrence of $h(x)$ with $\hat{h}(x)$:
$\widehat{r[h]} = r[\hat{h}]$. As for any linear observable, its vacuum
expectation value vanishes, $\langle r[\hat{h}] \rangle = 0$. The
emission time observable is then quantized perturbatively as
\begin{equation}\label{eqhsmdef}
	\hat{\tau}(s) = \tau_\mathrm{cl}(s)(1 + r[\hat{h}]) + \O(\hat{h}^2)
\end{equation}
and the variance of the emission time is
\begin{align}
	(\Delta\tau)^2
	&= \langle \hat{\tau}(s)^2 \rangle - \langle \hat{\tau}(s) \rangle^2 \\
\label{eqdtau}
	&= \tau_\mathrm{cl}(s)^2(1 + \langle r[\hat{h}]^2 \rangle) + \O(\ell_p^2) .
\end{align}

Unfortunately, as discussed in Sec.~VII~C of~\cite{khavkine}, the above na\"ive
expression for $(\Delta\tau)^2$ is divergent due to the $x\to y$
coincidence singularity on the right-hand side of
Eq.~\eqref{eqhadamarddef}. A physically motivated way of regularizing
this divergence is to recall that field measurements are, in any case,
never localized with infinite spacetime precision~\cite{bohr-rosenfeld,
bergmann-smith}. Thus, we are justified in replacing the point field
$\hat{h}(x)$ with the smeared field
\begin{equation}
	\tilde{h}(x)= \int \mathrm{d}z \, \hat{h}(x-z) \,  \tilde{g}(z) ,
\end{equation}
where $\tilde{g}(z)$ is the smearing function and can be interpreted as
the detector sensitivity profile. It phenomenologically models all
possible sources of smearing, including the fluctuations in the
center-of-mass positions of the lab and probe equipment, as well as the
finite spatial and temporal resolution of the signal emission and
reception. The expectation value $\langle r[\tilde{h}]^2 \rangle$ is
then finite, though dependent on some moments of the detector
sensitivity profile.  This observation simply shows that the quantum
aspects of the time delay observable depend on a few more details of the
lab and probe material models than just its purely classical aspects.

\section{Provisional choices}
\label{secchoices}
While the summary of Sec.~\ref{sectimedelay} make it clear how to go
about computing the quantum vacuum fluctuation in the time delay
observable, there remain several concrete choices to be made to fully
define the steps of such a calculation. These choices are discussed
explicitly below. Not all of these choices are ideal and should be
re-examined and improved in future work.

\subsection{Truncation order}
\label{subsectruncorder}
We are interested in computing the quantum vacuum fluctuation
$(\Delta\tau)^2$ given by Eq.~\ref{eqdtau}. We have an expression for
$\tau(s)$ valid to order $\O(h)$. So, upon quantization, we expect to
get an expression for $(\Delta\tau)^2$ valid to the same order. However,
at that order, the correction must be proportional to the expectation
value $\langle r[\hat{h}] \rangle$, which vanishes by virtue of being
linear in $\hat{h}$. Therefore, the leading nontrivial contribution
$(\Delta\tau)^2$ is of order $\O(\ell_p^2)$, where we have noted that,
after taking the vacuum expectation value, an operator correction of
order $O(\hat{h}^n)$ translates to a correction of order $\O(\ell_p^n)$
if $n$ is even and vanishes otherwise. To get a correct expression at
that order, we must know $\tau(s)$ to order $O(h^2)$ to begin with,
\begin{equation}
	\tau(s) = \tau_\mathrm{cl}(s)(1+r[h]+r_2[h]) + O(h^3) .
\end{equation}
Then
\begin{align}
	(\Delta\tau)^2
	&= \tau_\mathrm{cl}(s)(1+\langle r[\hat{h}]^2 \rangle + \langle
	\widehat{r_2[h]} \rangle) + \O(\ell_p^4) .
\end{align}

The quadratic correction $r_2[h]$ is partially
	\footnote{What is computed in Appendix~\ref{apppertsol} is the
	solution of the geodesic equation to order $\O(h^2)$ from which the
	$r_2[h]$ correction can be extracted.} %
computed in Appendix~\ref{apppertsol}. However, we do not include it in
the quantum vacuum fluctuation in this paper. The main reason is that of
feasibility. As will be seen in Sec.~\ref{seccalc}, the evaluation of
the $\langle r[\hat{h}]^2 \rangle$ (or rather its smeared version) is
already quite involved and the term $\langle \widehat{r_2[h]} \rangle$
would be even more complicated, as evidenced by the expressions given in
Appendixes~\ref{apppertsol} and~\ref{applinobsv}. Also, $r_2[h]$ does
not appear if we treat linearized gravity as an independent theory and
$r[h]$ a gauge invariant observable of independent interest. We adopt
this interpretation below. Thus, this result is a toy model for a result
that could be expected from the one involving $r_2[h]$, which itself
would be a toy model for the result of a higher perturbative order or
even nonperturbative calculation. Future work should incorporate the
quadratic $r_2[h]$ term directly into the calculation.

\subsection{Graviton two-point function}
The Wightman two-point function $\langle \hat{h}(x) \hat{h}(y) \rangle$
strongly depends on the choice of gauge. However, the expectation value
of any gauge invariant observable is independent of this choice. So we are
free to select, from the possible choices, a form of the two-point
function that is convenient for our purposes. In fact, we select it such
that the symmetrized (Hadamard) two-point function takes the simple and
covariant expression
\begin{gather}\label{eqhadamard}
	\langle \{ \hat{h}_{ij}(x) , \hat{h}_{kl}(y) \} \rangle
	= \frac{\ell_p^2}{\pi} P \frac{\eta_{ij,kl}}{(x-y)^2}, \\
\label{eqeta}
	\eta_{ij,kl} 
	= \eta_{ik}\eta_{jl} + \eta_{il}\eta_{jk} - \eta_{ij}\eta_{kl} ,
\end{gather}
where $P$ denotes a Cauchy principal value distribution. This formula is
justified in Appendix~\ref{apptwopoint}.
We are ultimately interested in computing the
vacuum fluctuation in the quantized emission time observable
$\hat{\tau}(s)$, which in our approximation reduces to computing the
expectation value of the square of the quantized linear correction
$r[\hat{h}]$. The latter quantity is expressible in terms of the
Hadamard two-point function~\eqref{eqhadamard}.

\subsection{Smearing profile}
\label{subsecsmearing}
Unfortunately, without a detailed model of the lab and probe equipment,
there is no natural choice for the smearing profile $\tilde{g}(x)$ in
the definition of the smeared graviton field $\tilde{h}(x)$ in
Eq.~\eqref{eqhsmdef}. We make the following pragmatic choice that
balances generality and simplicity in the resulting calculations
\begin{gather}
	\int \d{z} \, \tilde{g}(z) = 1 , \\
	\tilde{g}(z) = \bar{g}(z_\perp^2)\delta(u\cdot z) ,
\end{gather}
where $u$ is the unit vector parallel to the lab worldline, $z_\perp
= z + (z\cdot u) \, u$,  $z_\perp^2=R^2$, and $\bar{g}(R^2)$ is smooth and strongly peaked around $R=0$.
As will be seen below, the profile that will directly appear in the
results is rather the self-convolution
\begin{equation} \label{gprofile}
	\tilde{g} * \tilde{g}(z)
	= \int \d{x} \, \tilde{g}(z-x) \tilde{g}(x)
	= \frac{1}{4\pi} g(z_\perp^2) \delta(u\cdot z) ,
\end{equation}
where $g(R^2)$ has the same characterization as $\bar{g}(R^2)$. This
choice of $\tilde{g}(z)$ is simple, is invariant under rotations fixing
$u$, ensures that the self-convolution $\tilde{g} * \tilde{g}(z)$ is
equally simple and symmetric, and is still general enough to allow its
moments to be essentially arbitrary. We only require that there exists a
length scale $\mu$ (the \emph{smearing scale}) such that arbitrary
moments behave like
\begin{equation} \label{eqmoments}
	\int \d{z} \, z^k \, \tilde{g}(z) \sim \mu^k ,
\end{equation}
with coefficients of proportionality of order $\O(1)$.

\begin{figure}
\centering
\includegraphics[scale=1]{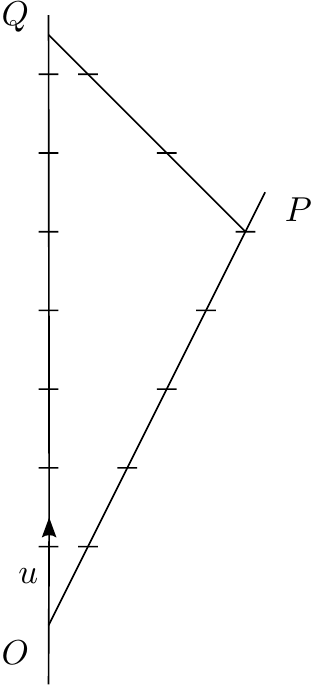} 
\caption{Rough sketch of the support of the smearing function
$\tilde{g}(z)$ overlaid on the $OPQ$ geodesic triangle.}
\label{figsmear}
\end{figure}

Unfortunately, this pragmatic choice explicitly breaks Lorentz
invariance. The effect of the smearing along the geodesic triangle is
illustrated in Fig.~\ref{figsmear}. The smearing profile $\tilde{g}(z)$
must break Lorentz symmetry in some way, otherwise it could not be
peaked only near $z=0$. However, it would be more physically
reasonable to suppose that the local geometry of each geodesic
determines the orientation of the smearing profile at its own points.
Unfortunately, that would reduce the symmetry of the cross-convolutions
of the different smearing profiles and hence significantly complicate
the estimation of their moments. Future work should deal with such
complications and use a more physically reasonable smearing scheme. We
hope, though, that the results would not be qualitatively significantly
different from the present work.

\section{Calculation}
\label{seccalc}
In this section, we describe the calculation of the quantum variance of
the time delay, the core of this paper, in more detail. The details are
presented in four parts. The first part, Sec.~\ref{subsecmaster},
derives a master formula for the quantum variance. This master formula
is based on the structure of linearized time delay observable
[Eqs.~\eqref{eqimplicitr} and~\eqref{eqexplicitr}] and encapsulates all
quantum expectation values in smeared segment integrals.  The
smeared segment integrals contain two kinds of integrations performed on
the graviton Hadamard two-point function: one-dimensional integrals over
background geodesic segments and four-dimensional integrals over a
smearing function. The segment integrations and the angular smearing
integrals are to be precalculated and tabulated as described in
Sec.~\ref{secgeneratingtables}, which constitutes the second part.
Section~\ref{secRTsmearing}, completes the description of the smeared
segment integrals. Finally, Sec.~\ref{secupdmaster} describes how these
tables can then be used to efficiently compute, using an updated master
formula, the quantum variance of the time delay for an arbitrary shape
of the corresponding geodesic triangle, and potentially for other
thought experiment geometries.

The algorithm described below was implemented using computer algebra
software (\textsc{Mathematica}~8.0). The results of the calculations
carried out with its help are described in Sec.~\ref{secresults}.

\subsection{Master formula for $\expr$}
\label{subsecmaster}
We denote the smeared first-order correction to the time delay as follows
\begin{equation}
	\tilde{r} = r[\tilde{h}] = \sum_{KmX} r^{Kij}_{mX} \int^{(m)}\!\d{z} \nabla_K \tilde{h}_{ij}(z)
\end{equation} 
and we write $\langle \tilde{r}^2 \rangle$ for the corresponding smeared
correction to the variance of the time delay. Below we derive a master
formula for this quantum variance that separates the geometric aspects
of the time delay observable, as encapsulated in the coefficients
$r^{Kij}_{mX}$, and the quantum effects, as encapsulated in the
\emph{smeared segment integrals} $\tilde{I}^{mn}_{K}(X,Y)$ to be
introduced below. The capital letter $K$ (and later $L$) denote
multi-indices (cf.~Sec.~\ref{subseclinearized}).

The quantum variance can be written as 
\begin{widetext}
\begin{align}
\notag	\langle\tilde{r}^2\rangle
			&= \frac{1}{2}\langle\{\tilde{r},\tilde{r}\}\rangle \\
\notag		&= \frac{1}{2}\sum_{KmX}\sum_{LnY} r^{Kij}_{mX} r^{Lkl}_{nY}
		\int\!\d{x}\int\!\d{y}\, \nabla_K\tilde{g}_u(x) \nabla_L\tilde{g}_u(y)
		\int_X^{(m)}\!\d{x'}\int_Y^{(n)}\!\d{y'}\,
			\langle\{\hat{h}_{(ij)}(x'-x),\hat{h}_{(kl)}(y'-y)\}\rangle \\
\notag		&= \frac{\ell_p^2}{2\pi}\sum_{KmX}\sum_{LnY}
			r^{Kij}_{mX}\,\eta_{ij,kl}\,r^{Lkl}_{nY}
		\int\!\d{x}\int\!\d{y}\, \nabla_K\tilde{g}_u(x) \nabla_L\tilde{g}_u(y)
		\int_{X-x}^{(m)}\!\d{x'}\int_{Y-y}^{(n)}\!\d{y'}\,P\frac{1}{[y'-x']^2} \\
\notag		&= \frac{\ell_p^2}{2\pi}\sum_{KmX}\sum_{LnY}
			r^{Kij}_{mX}\,\eta_{ij,kl}\,r^{Lkl}_{nY}
		\int\d{x}\int\d{y}\, \nabla_K\tilde{g}_u(x) \nabla_L\tilde{g}_u(y)
			I^{mn}(X-x,Y-y) \\
\notag		&= \frac{\ell_p^2}{2\pi}\sum_{KmX}\sum_{LnY}
			(-)^{|L|} r^{Kij}_{mX}\,\eta_{ij,kl}\,r^{Lkl}_{nY}
		\int\d{z}\,I^{mn}(X,Y+z)
			\int\d{y}\,\nabla_{K\cup L}\tilde{g}_u(z+y) \tilde{g}_u(y) \\
\notag		&= \frac{\ell_p^2}{2\pi}\sum_{KmX}\sum_{LnY}
			(-)^{|L|} r^{Kij}_{mX}\,\eta_{ij,kl}\,r^{Lkl}_{nY}
		\int\d{z}\,I^{mn}(X,Y+z) \nabla_{K\cup L} g_u(z) \\
		&= \frac{\ell_p^2}{2\pi}\sum_{KmX}\sum_{LnY}
			(-)^{|L|} r^{Kij}_{mX}\,\eta_{ij,kl}\,r^{Lkl}_{nY}
			\tilde{I}^{mn}_{K\cup L}(X,Y), \label{eqbegincalc}
\end{align}
\end{widetext}
where in the first line we used the Hadamard two-point
function~\eqref{eqhadamard} and we introduced the following definitions
\begin{align}
	I^{mn}(X,Y)
	&= \int^{(m)}_X\d{s}\int^{(n)}_Y\d{t}\, P\frac{1}{[x(s)-y(t)]^2} , \\
	\tilde{I}^{mn}_K(X,Y)
	&= \int\d{z}\,I^{mn}(X,Y+z) \nabla_K g_u(z), \label{eqitilde} \\
	g_u(z)
	&= \int\d{z'}\,\tilde{g}_u(z+z')\tilde{g}_u(z') \\
	&= (\tilde{g}_u*\tilde{g}_u)(z) . 
\end{align}
where ${*}$ denotes convolution [recall the relation
$\tilde{g}_u(-z')=\tilde{g}_u(z')$]. The convolved smearing function has
the same properties as the original smearing function as discussed in
Sec.~\ref{subsecsmearing}. Note the translation invariance
$I^{mn}(X+z,Y+z)=I^{mn}(X,Y)$. Even though the final  expression for
$\langle\tilde{r}^2\rangle$ does not appear to be symmetric under the
interchange of the $K$ and $L$ multi-indices, in fact, the extra
$(-)^{|L|}$ factor symmetrizes the interchange property
$\tilde{I}^{mn}_{K\cup L}(X,Y) = (-)^{|K|+|L|}\tilde{I}^{nm}_{L\cup
K}(Y,X)$, where $K\cup L = L\cup K$ is the concatenation of two
multi-indices.

The bulk of the work lies in evaluating the $\tilde{I}^{mn}_{K\cup
L}(X,Y)$ integral. Since for each term in $\tilde{r}^2$ we have such an
integral, and $\tilde{r}$ consists of ten terms, we have to evaluate
$\frac{1}{2}\cdot10 \cdot 11=55$ such integrals. Additionally, each
integral contains $6$--$8$ one-dimensional integrals, which makes a
total of ${\sim}400$ one-dimensional integrals. This is not the entire
story yet, looking closer at the integrals one notices that the
singularity structure changes depending on whether the line segments
along which the integral needs to be evaluated are either timelike or
null and parallel or nonparallel. Together with some additional
technical details to be discussed, this results in ten different
singularity structures.

In short, there is no simple, direct master formula that can be given
for the evaluation of (the leading $\mu$-order expansion terms of) the
smeared segment integrals $\tilde{I}^{mn}_{K}(X,Y)$. Instead, we settle
for the master formula~\eqref{eqwithunderbrace} of intermediate type.
Part of it can be evaluated symbolically and tabulated for different
argument types. The remaining part can be evaluated numerically as
needed using an algorithm with table look-ups. All these (hybrid
numerical-symbolic) operations are automated using the computer algebra
software \textsc{Mathematica}~8.0. The details of each of the two parts of
the calculation are discussed below.

\subsubsection{Spherical coordinates for smearing}
\label{subsecazimuthal}
The $\tilde{I}^{mn}_{K}(X,Y)$ integral is completely determined by the
number of derivatives $|K|$ on the smearing function  $(|K|=0,1,2)$, the
number of iterated integrals along the $X$ and $Y$ segments denoted by
$m$ and $n$ (where $m=0,1$ and similarly for $n$) and the line segments
along which the integrals need to be evaluated. We decompose $z=
-(u\cdot z)z + (\u \cdot z) \u + w$, where $\u$ is a spacelike unit
vector, taken to be $\u^i=(0,1,0,0)$  (hence $u \cdot \u=0$) and $w$ is
orthogonal to the $(u,\u)-$plane. We parametrize $z$ as
\begin{align}
	z &=\left(- u \cdot z, \u \cdot z, w^1,w^2 \right) \\
		&= \left( T, R \cos \theta, R \sin \theta \cos \phi,R \sin \theta \sin \phi \right)
\end{align}
and write the four-dimensional integral over the spacetime separation
$z$ in $\tilde{I}^{mn}(X,Y+z)$ as
\begin{align}
\int\d^4{z} &= \int\!\d{(u\cdot z)} \int\!\d{(\hat{u}\cdot z)} \int\!\d^2{w} \\
&= \int_{-\infty}^{\infty}\!\d{T} \int_{-R}^{R}\!\d{c} \int_0^{2\pi}\!\d{\phi} \int_0^\infty\!\d{R} \, R,
\end{align}
where we defined $c=R \cos \theta$, with $w^2= R^2 - c^2$.

As discussed in Sec.~\ref{subsecsmearing}, the smearing function is set
to $g_u(z)=g(z_\perp^2)\delta(u\cdot z)$. The smearing function with any
number of derivatives can be written compactly as
\begin{equation} \label{derg}
	\nabla_K \, g_u(z) = \sum_{\T,d,\gamma,p,l}
		\T_K \delta^{(d)}(-T) \, g^{(\gamma)}(R^2) \,  R^{p} \, c^l \,
		P_{\T,d,\gamma,p,l}  ,
\end{equation}
where $K$ is a multi-index, $P_{\T,d,\gamma,p,l}$ are numerical
coefficients, $d,\gamma,p,l$ range over a non-negative finite integral
set, and $\T_K$ ranges over a certain basis of rank-$|K|$ tensors
consisting of symmetrized products of $u$, $\hat{u}$ and $\delta_{\perp}
= \eta + u u$. The coefficients are nonzero only when the indices
satisfy the homogeneity constraint $d+2\gamma-p=|K|$. For the integrals
we are considering, the maximal number of derivatives on the smearing
function is two. Then, $\T_K$ ranges over either $\{1\}$ for $|K|=0$,
$\{u, \hat{u}\}$ for $|K|=1$, or $\{uu, \hat{u}\hat{u},
u\hat{u}+\hat{u}u, \delta_{\perp}\}$ for $|K|=2$.  The maximal power of
$c$ in $P_{\T,d,\gamma,p,l}$ is also two.  The exact expression for all
the required derivatives of the smearing function can be found in
Table~\ref{tablesmearing}.

Since the smearing function is independent of the direction of $w$ and
$I^{mn}(X,Y)$ depends only on $w^2$, $g_u(z)$ and its derivatives can be
independently integrated (or \emph{averaged}) over the directions of
$w$. The averaging procedure for $w$ is fairly straightforward. For
symmetry reasons, all terms that are odd in $w$ when averaged give zero.
Looking at the second column of Table~\ref{tablesmearing}, we also need
the following integral identities (where we take $\hat{w}^2 = 1$ and $w =
\sqrt{R^2-c^2} \hat{w}$):
\begin{align}
\label{eqwavg1}
	\frac{1}{2\pi} \int \d^2{\hat{w}} \,
		&= 1  , \\
\label{eqwavg2}
	\frac{1}{2\pi} \int \d^2{\hat{w}} \, w_i w_j
		&= \frac{1}{2} (R^2-c^2) (\delta^\perp_{ij}-\u_i\u_j) ,
\end{align}
where $\delta^\perp_{ij} = \eta_{ij} + u_i u_j$ and the integration is
over a unit sphere, the possible values of $\hat{w}$. The tensor
structure of the last identity follows directly from the rotational and
reflection invariance of the integral, with the overall constant fixed
by computing its trace.

\begin{table*}
\centering
\begin{ruledtabular}
\begin{tabular}{p{5em}ll}
	& chain rule & $w$-averaging \\
\hline \\[-1ex]
$g_u(z)$
	& $ g(z_\perp^2) \delta(u\cdot z) $
	& $ g(R^2)\delta(-T)$ \\ 
$ \nabla g_u(z) $
	& $ u g(z_\perp^2) \delta'(u\cdot z)
		+ 2[(\u\cdot z)\u+w] g'(z_\perp^2) \delta(u\cdot z) $
	& $ u g(R^2) \delta'(-T) + 2\u c g'(R^2) \delta(-T)$ \\		
$ \nabla \nabla g_u(z) $
	& $ uu g(z_\perp^2) \delta''(u\cdot z)
		+ 4 [u\u(\u\cdot z)+uw] g'(z_\perp^2) \delta'(u\cdot z)$ 
	& $ uu g(R^2) \delta''(-T) + 4 u \u c g'(R^2) \delta'(-T) $ 
	\\
	&
	$ \quad {}
		+ [2 \delta_\perp g'(z_\perp^2) $ 
	&
	$ \quad {}
		+ [2 \delta_\perp g'(R^2) $ 
	\\
	&
			$\qquad {}
			+ 4 (\u\u(\u\cdot z)^2
				+ 2\u w (\u\cdot z) + w w) g''(z_\perp^2)]  \delta(u\cdot z) $
	&
			$\qquad {}
			+ (4 \u\u c^2
				+ 2(\delta_\perp-\u\u)(R^2-c^2)) g''(R^2)] \delta(-T) $ \\
\end{tabular}
\end{ruledtabular}
\caption{
	Smearing function $g_u(z) = g(z_\perp^2) \delta(u\cdot z)$
	[cf.~Eq.~\eqref{gprofile}], with zero, one or two derivatives.  The
	second column shows the derivative chain rule applied to the profile
	ansatz. The third column shows the result after $w$-averaging, as
	discussed in Sec.~\ref{subsecazimuthal} and expressed in
	$(R,T,c=R\cos\theta)$ coordinates.  Products of vectors denote the
	symmetrized tensor product, e.g.,\ $(u w)_{ij} = u_{(i} w_{j)}$.
	Primes denote derivatives with respect to the argument of the
	corresponding function.
}
\label{tablesmearing}
\end{table*}

\subsubsection{Master formula for $\tilde{I}^{mn}_K(X,Y)$}
\label{subsecsegtint}
Substitution of the differentiated smearing function~\eqref{derg} into
the definition~\eqref{eqitilde} of $\tilde{I}^{mn}_K(X,Y)$ and recalling
that $\d^4 z = \d{c}\,R\d{R}\,\d{T}\,\d\phi$ gives
\begin{multline} \label{eqwithunderbrace}
	\tilde{I}^{mn}_K(X,Y) = \sum_{\T,d,\gamma,p,l} \T_K P_{\T,d,\gamma,p,l} \\
	 \times \int_0^{2\pi}\! \d{\phi} \underbrace{
		\int_{-\infty}^{\infty}\!\d{T} \delta^{(d)}(-T)
		\int_0^\infty\!\d{R} \, R^{p+1} g^{(\gamma)}(R^2)}_{\text{part II}
		} \\
	 \times
		\underbrace{
		\int^{(m)}_{X}\!\d{s} \int^{(n)}_Y\!\d{t} \int_{-R}^{R}\!\d{c} \,
			P \frac{c^l}{(y(t)-x(s)+z)^2}}_{\text{part I}
		} \\
	= \frac{2\pi}{\mu^2} \sum_{i=0}^\oo \sum_\T \T_K \, \mu^i \,
		I^{mn}_{\T,i}(\ln\mu;X,Y) .
\end{multline}
Since the smearing function $g_u(z)$ depends only on $R$ and $T$, the
evaluation of this integral can be broken down into two parts: symbolic
evaluation and tabulation (indicated by ``part~I'') after which the
remaining smearing can be performed (indicated by ``part~II'').  Note
that the $\phi$ integral simply results in the overall factor of $2\pi$
displayed on the last line of~\eqref{eqwithunderbrace}. Note that,
because we do not assume a precise form of the smearing function, we are
also not interested in an exact answer for $\tilde{I}^{mn}_K(X,Y)$.
Instead, as indicated above, we are only interested in a few of its
leading-order terms in the limit of small smearing scale $\mu$
[cf.~Eq.~\eqref{eqmoments}], namely the coefficients
$I^{mn}_{\T,i}(\ln\mu;X,Y)$ for small values of $i$. The dependence on
$\ln\mu$ in $I^{mn}_{\T,i}(\ln\mu;X,Y)$ is expected to be a low-order
polynomial.  Therefore, we take the opportunity to simplify the
calculations in ``part~I'' by judiciously expanding some of the
intermediate results in powers of $R$ and $T$ (also with logarithmic
terms, where appropriate).

\subsection{Tabulating angular and segment integrals} 
\label{secgeneratingtables}
In this section, we focus on evaluating the segment integration and the
remaining angular integration of the smeared segment integrals
$\tilde{I}^{mn}_K$, that is, ``part~I'' of~\eqref{eqwithunderbrace}. These
integrals can be evaluated analytically and for any given parameters (to
be specified below). Thus, they can be tabulated in advance for the
values of the parameters needed to compute $\expr$, even before the
triangular geometry is specified. This flexibility is what allows our
methods to be straightforwardly extended to observables with more
general underlying geometries.

We evaluate integrals of the following form, parametrized by integers
$m$, $n$ and $l$:
\begin{equation}\label{cstint}
	I^{mn}_l(R,T;X,Y)
		= \int^{(m)}_X\d{s} \int^{(n)}_Y\d{t}
			\int_{-R}^{R}\d{c} \,P \frac{c^l}{z(s,t,c)^2},
\end{equation}
where we will need $l=0,1,2$ and $m,n=0,1$. 

The integration is carried out in several steps. Note that we start with
a rational expression in all variables ($c$, $s$, $t$, $X$ and $Y$
endpoint coordinates). The integration with respect to $c$ is carried
out in Sec.~\ref{subseccint} and turns it into a mix of rational and
logarithmic terms, with a precisely controlled structure. Next, the $s$
and $t$ integrals are considered. If the segments $X$ and $Y$ are
nonparallel, it is advantageous to change coordinates
(Sec.~\ref{subsecnonprl}) to simplify the denominators and the
logarithmic arguments and then apply Stokes' theorem to convert the
two-dimensional integral into a one-dimensional one. A similar goal is
achieved for parallel segments using an alternative method
(Sec.~\ref{subsecprl}). In either case, iterated integrals are converted
to noniterated ones. The results for both the parallel and nonparallel
cases fit into the same precisely controlled structure, involving
rational functions and logarithms, which is fed into the following step.
The remaining one-dimensional integrals are evaluated
(Sec.~\ref{subsecparamseg}) and the result is a mix of rational,
logarithmic and dilogarithmic terms, again with a precisely controlled
structure.

At this stage, we will have an algorithm to compute explicit, exact
expressions for the integrals $I^{mn}_l(R,T;X,Y)$ defined
in Eq.~\eqref{cstint}, even when the coordinates of the endpoints of $X$ and
$Y$ are given symbolically. The only caveat is that cases when $X$ and
$Y$ are or are not parallel must be distinguished by hand. However, it
is not these expressions that we need, but their smeared derivatives
$\tilde{I}^{mn}_K(X,Y)$ or, even more precisely, the expansion
coefficients $\tilde{I}^{mn}_{\T,i}(\ln\mu;X,Y)$ defined in
Eq.~\eqref{eqwithunderbrace}. Note that the smeared
segment integrals $\tilde{I}^{mn}_K(X,Y)$ have singular leading terms in
the $\mu$ expansion only if the $X$, $Y$ segments have common or
lightlike separated endpoints. (All of these possibilities occur in the
time delay geometry.) These $\mu$-singularities stem from the singular
behavior of $I^{mn}_l(R,T;X,Y)$ for small $R$ and $T$ under the same
circumstances. Unfortunately, the structure of the $R,T$ singularities
depends strongly on more details of the relative geometry of the $X$ and
$Y$ segments. The $R,T$ expansion is performed and tabulated for each of
the possible cases (see Sec.~\ref{subsecsing} and Fig.~\ref{figdecision}).

These tables serve as input to ``part~II'', the remaining $R,T$ smearing
(Sec.~\ref{secRTsmearing}), which ultimately computes the
$\tilde{I}^{mn}_{\T,i}(\ln\mu;X,Y)$ coefficients.

\subsubsection{Integration with respect to $c$}
\label{subseccint}
When we confine the line segments and the displacement due to smearing
to the $(u,\u)-$plane, the denominator in~\eqref{cstint} can be
rewritten with the following notation:
\begin{align}
z(s,t,c)
	&= y(t) - x(s) + Tu+c\u+w , \\
z(s,t,c)^2
	&= -z_0^2+z_1^2 + 2cz_1 + R^2 , \\ 
\label{eqz0withT}
z_0 &= -u\cdot [y(t)-x(s)+Tu] , \\
\label{eqz1withT}
z_1 &= \u\cdot [y(t)-x(s)+Tu] ,
\end{align}
where we have obviously separated the $Tu$ and $c\u$ smearing shifts.
In this form, we see that the denominator depends only linearly on $c$,
which makes integration with respect to $c$ rather straightforward.
Basically, the integral consists of logarithms with the denominator
evaluated at $c=\pm R$ as arguments. This result simplifies even more
since the arguments of the logarithms factor as follows:
\begin{align}
	z(s,t,c=\pm R)^2
		&= -z_0^2 + (z_1+c)^2 \\
		&= (c+z_1-z_0)(c+z_1+z_0) \\
		&= z_{c+} z_{c-} , \\
	z_{c\pm} &= c+z_1\mp z_0=c+z_\pm ,
\end{align}
where we have introduced the new notation $v_\pm = v\cdot(\u\pm u)$ for
any vector $v$. After the $c$ integral has been performed, the symbol
$c$ will always refer to the possible endpoint values $\pm R$.

Performing the integration over $c$ in terms of these new variables
$\zcpm$ and $z_1$ yields
\begin{widetext}
\begin{align}
\int_{-R}^{R}\d{c} \frac{c^l}{-z_0^2+z_1^2 + 2cz_1 + R^2} &= \sum_{c=\pm R} \pm  \left( 2 \bar{P}_1(c,z_0;z_1) + P_2(z_{c\pm};z_1) [\ln|z_{c+}| + \ln|z_{c-}|] \right) \\
&= \sum_{c=\pm R} \pm \sum_\pm \left( P_1(c,z_{c\pm};z_1) + P_2(z_{c\pm};z_1) \ln|z_{c\pm}| \right),
\label{eqcint}
\end{align}
\end{widetext}
where the $\pm$-symbol following the summation over $c$ matches the sign
in this summation. The terms $\bar{P}_1, P_1$ and $P_2$ are polynomials
in the arguments before the semi-colon and Laurent polynomial in the
arguments after the semi-colon. The first two are related by
\begin{equation}
	P_1(c,z_{c\pm};z_1) = \bar{P}_1(c,\mp(z_{c\pm}-z_1-c);z_1) ,
\end{equation}
since expression $z_0$ in terms of $z_1$ and $z_{c\pm}$ in this way
allows to introduce an overall $\pm$-sum. Since $z_1$ appears
Laurent polynomially, the individual summands in the result of the $c$
integral may have poles for $z_1=0$. However, the integral we started
with was regular for $z_1=0$ and thus these singularities need to vanish
in the final result. This served as a consistency check on our
calculations (Secs.~\ref{subsecparamseg} and~\ref{secchecks}).

Next, integration over $s$ and $t$ must be performed. This is done in
different ways for the case when $X$ and $Y$ are parallel or
nonparallel segments.

\subsubsection{Variable change for nonparallel line segments}
\label{subsecnonprl}
\begin{figure}[h]
\includegraphics[scale=1]{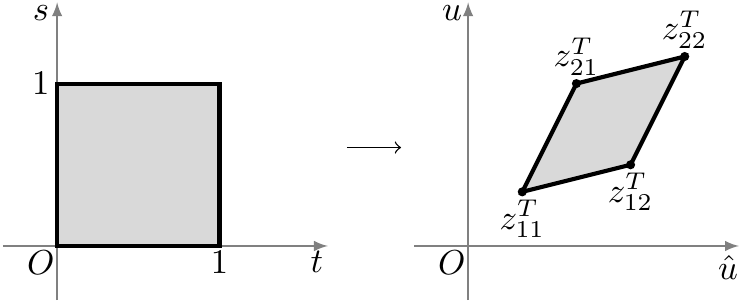}
\caption{Illustration of the change of variables from the $(s,t)$- to
	the $(u,\u)$-plane. We denote $z^T_{\mu\nu} = z_{\mu\nu}+Tu$,
	cf.~\eqref{eqzijdef}.}
\label{figchangeofvar}
\end{figure}

We can trade the complexity of iterated $s$ and $t$ integrals for
increased complexity of the integrands. The iterated integrals can be
treated similarly as the single integrals using Cauchy's formula
\begin{equation}
	\int_X^{(m)}\d{s} = \int_X\d{s}\,\frac{(1-s)^m}{m!}.
\end{equation}
At this point, we note that the integrands depend on $s$ and $t$
explicitly and through the expressions $\zcpm$ and $z_1$. If the $X$ and
$Y$ segments are nonparallel, the latter two are linearly independent
and thus can serve as alternative integration variables to $s$ and $t$.
It turns out to be advantageous to use $\zcpm$ and $z_1$ as the basic
integration coordinates, with the integration domain being the
parallelogram in the $(u,\u)$-plane spanned by the vector
$y(t)-x(s)+Tu$. This change of variables and the new integration domain
are illustrated in Fig.~\ref{figchangeofvar}, where we use the notation
\begin{align}
\label{eqzijdef}
z_{\mu\nu} &= y_\nu - x_\mu , ~~ \rlap{$x = x_2-x_1, ~ y = y_2-y_1 ,$} \\
y(t) &= y_1 + (y_2 - y_1) t , & y(0)&=y_1, \, y(1)=y_2 , \\
x(s) &= x_1 + (x_2 - x_1) s , & x(0)&=x_1, \, x(1)=x_2 .
\end{align} 

The explicit change of variables is
\begin{gather}
	s = \frac{y\wedge(z-z_{11})}{x\wedge y} ,
		\quad
	t = \frac{x\wedge(z-z_{11})}{x\wedge y} , \\
	\d{s}\wedge\d{t}
		= -\frac{\d{z_0}\wedge\d{z_1}}{x\wedge y}
		= \pm\frac{\d{z_{c\pm}}\wedge\d{z_1}}{x\wedge y} ,
\end{gather}
where we have used the following $\wedge$ notation and identity between
vectors (though, note that $\d{s}\wedge \d{t}$ stands for the usual
wedge product of differential forms):
\begin{align}
	v\wedge w &= - (v \cdot u) (w \cdot \u) + (w \cdot u) (v \cdot \u) , \\
	v\wedge w &= \pm[(v\cdot\u)w_\pm-(w\cdot\u)v_\pm] . 
\end{align}
Clearly, this transformation becomes singular when $X$ and $Y$ are
parallel ($x\wedge y = 0$). That case is handled differently in the next
subsection.

So the integral we are interested in is
\begin{widetext}
\begin{align} 
\MoveEqLeft
\int_X^{(m)}\!\d{s} \int_Y^{(n)}\!\d{t} \left( P_1(z_{c\pm};z_1) + P_2(z_{c\pm};z_1) \ln \vert z_{c\pm} \vert \right) \\
&= \int_0^1\!\d{s} \int_0^1\!\d{t} \frac{(1-s)^m}{m!} \frac{(1-t)^n}{n!} \left( P_1(z_{c\pm};z_1) + P_2(z_{c\pm};z_1) \ln \vert z_{c\pm} \vert \right) \\
&= \pm \int_{X\wedge Y} \frac{\left(1 - \frac{y\wedge(z-z_{11})}{x\wedge y}\right)^m}{m!} \frac{\left(1 - \frac{x\wedge(z-z_{11})}{x\wedge y}\right)^n}{n!} \left( P_1(z_{c\pm};z_1) + P_2(z_{c\pm};z_1) \ln \vert z_{c\pm} \vert \right) \frac{\d{\zcpm} \wedge \d{z_1}}{x \wedge y}. \label{eqintbeforestokes}
\end{align}
\end{widetext}

Since any 2-form is closed (being top dimensional) by the Poincar\'e
lemma, it is also exact; i.e.,\ we can write the differentials 
in~\eqref{eqintbeforestokes} as $\d{Q}$ where $Q$ is some 1-form. Then by
Stokes's theorem, we can reduce the integral in~\eqref{eqintbeforestokes}
from an integral over the interior to an integral over the boundary of
the parallelogram.  The boundary of the $st$ integration domain is
$(0,0)\stackrel{s}{\longrightarrow}
 (1,0)\stackrel{t}{\longrightarrow}
 (1,1)\stackrel{-s}{\longrightarrow}
 (0,1)\stackrel{-t}{\longrightarrow}
 (0,0)$.
In terms of the $z$ integration domain, this becomes
$z_{11}\stackrel{-x}{\longrightarrow}
 z_{21}\stackrel{y}{\longrightarrow}
 z_{22}\stackrel{x}{\longrightarrow}
 z_{12}\stackrel{-y}{\longrightarrow}
 z_{11}$.
We can formalize this procedure as follows. We are to integrate an
expression of the form $P=\d Q$ where $Q$ is some 1-form. If we pull $Q$
back to any line, say $x$, then $Q_x$ is also top dimensional and
therefore closed.  Thus, we can write $Q_x =\d L_x$ with $L_x$ a 0-form,
so that
\begin{align}
\label{eqQsegint}
	\int_{X\wedge Y}P
	&= \int_{\del(X\wedge Y)} Q \\
\label{eqLsum}
	&= \sum_{\mu,\nu=1,2} (-)^{\mu+\nu} [L_y(z_{\mu\nu})-L_x(z_{\mu\nu})] ,
\end{align}
where $L_x$ corresponds to integration over an edge parallel to $x$ and
similarly for $L_y$. In short, by applying Stokes's theorem, we reduced
the two-dimensional integral over the interior of the parallelogram to
one-dimensional integrals over the edges. Moreover, these
one-dimensional integrals are reduced to a sum over their end points,
which are the vertices of the original parallelogram.

To actually get $Q$ from $P$, which is just the integrand in
Eq.~\eqref{eqintbeforestokes}, we simply perform the
$\zcpm$ integration. Under this operation, the structure of the
expression does not change:
\begin{multline} \label{eqzcpmint}
	\int\d\zcpm\, (P_1(\zcpm;z_1)+P_2(\zcpm;z_1)\ln|\zcpm|) \\
	= P_3(\zcpm;z_1) + P_4(\zcpm;z_1)\ln|\zcpm| .
\end{multline}
The reason is that the $P_i(\zcpm;z_1)$ coefficients depend polynomially
on $\zcpm$. The integration can then be done by elementary methods. The 
structure of $L$, obtained from $Q$, will be more complicated. It is
discussed in Sec.~\ref{subsecparamseg}.

\subsubsection{Variable change for parallel line segments}
\label{subsecprl}
As mentioned before, when the line segments are parallel, it is no
longer possible to construct an invertible transformation between
$(s,t)$ and $(z_0,z_1)$. This is easily seen from the fact that
parallelogram on the right of Fig.~\ref{figchangeofvar} collapses to a
segment. Unfortunately, also, starting with formulas for the
nonparallel case and taking a limit produces many technical
difficulties. We found that it is most convenient to treat the $s$ and
$t$ integrals in the parallel case separately, as is discussed below.

The failure to invertibly transform from $(s,t)$ to $(z_0,z_1)$
coordinates indicates that we can write $z_0$, $z_1$, $z(s,t,c)^2$ or
any function $F(z_0,z_1)$ as a function $F(\zeta(s,t))$ of some single
affine-linear combination $\zeta(s,t)$ of $s,t$ with nonzero constants
$\zeta_s=\d\zeta/\d{s}$ and $\zeta_t=\d\zeta/\d{t}$
	\footnote{The constant $\zeta_s$ or $\zeta_t$ would vanish only if one
	of the segments, $x$ or $y$, were of zero length. We are excluding
	this possibility.}. %
Each $s$ or $t$ integral can then be converted into a $\zeta$ integral. 
The iterated integrals are now handled recursively. Denote 
$\zeta=\zeta(s,t)$, $\zeta^S=\zeta(0,t)$, $\zeta^T=\zeta(s,0)$ and 
$\zeta^{ST}=\zeta(0,0)$ and define
\begin{align}
F_{m,n}
	&= \frac{F^{[m+n+2]}(\zeta)}{\zeta_s^{m+1}\zeta_t^{n+1}} 
		+ \sum_{k=0}^{m+n+2} p^S_{m,n;k}(s)\frac{F^{[k]}(\zeta^S)}{\zeta_t^{n+1}} \notag \\
& \quad{} + \sum_{k=0}^{m+n+2} p^T_{m,n;k}(t)\frac{F^{[k]}(\zeta^T)}{\zeta_s^{m+1}}	\notag \\
& \quad{} + \sum_{k=0}^{m+n+2} p^{ST}_{m,n;k}(s,t)F^{[k]}(\zeta^{ST}), \label{eqwithf}
\end{align}
where the $p$'s (to be defined below) are polynomials in their
arguments, while $F_{m,n} = F_{m,n}(\zeta,s,t)$ and
\begin{align}
\label{eqFsegint}
	\frac{\d}{\d\zeta} F^{[k+1]} &= F^{[k]}, \\
\notag
	F^{[0]} &= F(\zeta), \\
\notag
	F_{-1,-1} &= F(\zeta).
\end{align}
The structure of this expression is preserved under integrations with
respect to $s$ and $t$, with only the $p$'s changing, if we define
\begin{equation}
	\int_0\d{s}\,F_{m,n} = F_{m+1,n} \quad\text{and}\quad
	\int_0\d{t}\,F_{m,n} = F_{m,n+1}.
\end{equation}
Setting $s=t=1$ in $F_{m,n}$ precisely yields $I^{mn}_l$ defined in
Eq.~\eqref{cstint} for a proper choice of $F(\zeta)$. This choice is
just the result of the $c$ integration given in Eq.~\eqref{eqcint}, with
the replacements $z_1=z_1(\zeta)$ and $\zcpm=\zcpm(\zeta)$.

Note that the integration constants are chosen such that $F_{m,n}=0$
whenever either $s=0$ or $t=0$ for any $m\ge0$ or $n\ge0$. With the
above initial conditions, the polynomial coefficients will satisfy the
following recurrence relations:
\begin{align}
	p^S_{m+1,n;k}(s)
	&= \int_0\d{s}\,p^S_{m,n;k}(s) - \frac{\delta_{k,m+1+n+2}}{\zeta_s^{m+2}}, \\
	p^S_{m,n+1;k}(s)
	&= p^S_{m,n;k-1}(s), \\
	p^T_{m,n+1;k}(t)
	&= \int_0\d{t}\,p^T_{m,n;k}(t) - \frac{\delta_{k,m+n+1+2}}{\zeta_t^{n+2}}, \\
	p^T_{m+1,n;k}(t)
	&= p^T_{m,n;k-1}(t), \\
	p^{ST}_{m+1,n;k}(s,t)
	&= \int_0\d{s}\,p^{ST}_{m,n;k}(s,t) - \frac{p^T_{m,n;k-1}(t)}{\zeta_s^{m+2}}, \\
	p^{ST}_{m,n+1;k}(s,t)
	&= \int_0\d{t}\,p^{ST}_{m,n;k}(s,t) - \frac{p^S_{m,n;k-1}(s)}{\zeta_t^{n+2}}.
\end{align}
The coefficients that are relevant for the integrals we consider can be
found in Table~\ref{tabpcoeff}.

\begin{table}[h]
\centering
\begin{ruledtabular}
\begin{tabular}{cccccc}
\multicolumn{1}{l}{ }  & & $p$ & $p^S$ & $p^T$ & $p^{ST}$ \rule[1ex]{0pt}{1.5ex}\\
\hline
\rule[1ex]{0pt}{1.5ex} $m=0,n=0$ & $k=2$ & $\frac{1}{\zs \zt}$ & $-\frac{1}{\zs \zt}$ & $-\frac{1}{\zs \zt}$  & $\frac{1}{\zs \zt}$  \\
\rule[1ex]{0pt}{1.5ex} $m=1,n=0$ & $k=2$ &  & $-\frac{s}{\zs \zt}$ &  & $\frac{s}{\zs \zt}$ \\
\rule[1ex]{0pt}{1.5ex}
 & $k=3$ & $\frac{1}{\zs^2 \zt}$ & $-\frac{1}{\zs^2 \zt}$ & $-\frac{1}{\zs^2 \zt}$  & $\frac{1}{\zs^2 \zt}$ \\
\rule[1ex]{0pt}{1.5ex}
$m=0,n=1$ & $k=2$ &  &  & $-\frac{t}{\zs \zt}$ & $\frac{t}{\zs \zt}$ \\
\rule[1ex]{0pt}{1.5ex}
 & $k=3$ & $\frac{1}{\zs \zt^2}$ & $-\frac{1}{\zs \zt^2}$ & $-\frac{1}{\zs \zt^2}$  & $\frac{1}{\zs \zt^2}$ \\
\rule[1ex]{0pt}{1.5ex}
$m=1,n=1$ & $k=2$ &  &  &  & $\frac{s t}{\zs \zt}$ \\
\rule[1ex]{0pt}{1.5ex}
 & $k=3$ &  & $-\frac{s}{\zs \zt^2}$ & $-\frac{t}{\zs^2 \zt}$  & $\frac{s \zs + t \zt}{\zs^2 \zt^2}$ \\ 
 \rule[1ex]{0pt}{1.5ex}
 & $k=4$ & $\frac{1}{\zs^2 \zt^2}$ & $ -\frac{1}{\zs^2 \zt^2}$ & $-\frac{1}{\zs^2 \zt^2}$ & $\frac{1}{\zs^2 \zt^2}$ \\
\end{tabular}
\end{ruledtabular}
\caption{The polynomial coefficients from Eq.~\eqref{eqwithf} for the parallel case for different values of $m,n$ and $k$.}
\label{tabpcoeff}
\end{table}

Thus, also for the parallel situation, we are left to evaluate
one-dimensional integrals, in particular, integrals parametrized by
$\zeta(s,t)$. Our calculations require two-, three- and maximally four-iterated
integrals. One can think of these integrals in a similar way as for the
integrals in the nonparallel situation: the $\zeta(s,t)$ parametrizes
the sides of the parallelogram and the four terms in~\eqref{eqwithf}
correspond to the four edges of the parallelogram. 

In sum, for both situations, nonparallel and parallel line segments, we
are left to evaluate one-dimensional integrals along the sides of a
parallelogram. Evaluation of these one-dimensional integrals is
discussed next. 

\subsubsection{Edge segment integrals}
\label{subsecparamseg}
In the two preceding sections, we have converted the two-dimensional
integrals over $s$ and $t$ into one-dimensional integrals over the
boundary edges of the parallelogram on the right of
Fig.~\ref{figchangeofvar}. In the nonparallel case, these are the
integrals on the right-hand side of Eq.~\eqref{eqQsegint}. In the
parallel case, these are the integrals that solve Eq.~\eqref{eqFsegint}.
In either case, we need to find a convenient way to parametrize the edge
segments (we will use a parameter $\sigma$) and keep track of the
structure of the integrand. We address this below.

We again need to consider two different situations: one in which the
edge is completely in the direction of $u$ and one in which the edge
also has a $\u$ component. A different parametrization is needed for
each case. However, in both cases, each side of the parallelogram is
described by its starting point $b$ and its tangent vector $a$, which
runs from one vertex to the next. First the procedure for the latter
situation, which corresponds to  $a \cdot \u \neq 0$ is outlined and
successively the situation in which the edge is entirely in the
$u$ direction, that is, $a \cdot \u = 0$. 

\paragraph{Case $a\cdot\u\neq 0$.}
When $a \cdot \u \neq 0$, we parametrize each edge by $z(\sigma) =
z_0(\sigma) u + z_1(\sigma) \u$ with 
\begin{align}
z_0 (\sigma) &= B_0 - C_0 \sigma, \\
z_1 (\sigma) &= \sigma.
\end{align}
To relate the constants $B_0$ and $C_0$ to the geometry of the
parallelogram, we look at the ``velocity'' of the edge
\[
\frac{\d}{\d{\sigma}} z(\sigma) = K a,
\]
where $K$ is an unknown constant. If we dot this equation with $- u$ and
$\u$, we can compare this to the derivatives of $z_0$ and $z_1$ to
determine $C_0$ in terms of the $a$ and $b$ vectors
\[
\frac{-K (a\cdot u)}{K (a \cdot \u)} =\frac{\frac{\d{z_0}(\sigma)}{\d{\sigma}}}{\frac{\d{z_1}(\sigma)}{\d{\sigma}}}=\frac{-C_0}{1} \Longrightarrow C_0=\frac{a \cdot u}{a \cdot \u}.
\]
To determine $B_0$ in terms of the $a$ and $b$ vectors, we look at the
starting point of the edge which corresponds to $\sigma=0$. At this
point $z(\sigma=0) = b$, but also $z(\sigma=0)= z_0(0) u + z_1(0) \u$,
which upon applying $a\wedge u = (a\cdot \u) \, (\u \wedge u) = - a\cdot
\u$ shows that $B_0= - \frac{a \wedge b}{a \cdot \u}$. After integration
along the vertices, the start and end point of each segment needs to be
inserted, which is at each vertex $\sigma=\bdh$.

The ($\zcpm$,$z_1$) variables are related to these new variables as
follows. We already know that $z_1=\sigma$ and $\zcpm$ is obtained by
\begin{align}
\zcpm &= c + z_1 \mp z_0 \\
&= c + \sigma \mp B_0 \pm C_0 \sigma \\
&= \bpm \left( 1 - \cpm \sigma \right),
\end{align}
where we defined $\bpm=c \mp B_0$ and $\cpm = - \frac{1 \pm C_0}{c \mp
B_0}$. Hitherto, the shift in the $u$ direction from the temporal
smearing [see Eqs.~\eqref{eqz0withT} and~\eqref{eqz1withT}] has not been
explicitly taken into account. Fortunately, it can be simply re-obtained
by absorbing the shift in the $b$ vector: $\bdu \to \bdu - T$. This
gives
\begin{equation}
B_0 = - \frac{a \wedge b}{a \cdot \u} \longrightarrow - \frac{a \wedge b}{a \cdot \u} + T.
\end{equation}
$C_0$ does not change as it does not contain $b$. Thus, taking the shift
by the smearing into account, we have
\begin{align}
\bpm &=  \frac{(a \cdot \u) (c \mp T) \pm a \wedge b}{a \cdot \u}, \\
\cpm &= - \frac{a_{\pm}}{a \cdot \u (c \mp T) \pm a \wedge b}.
\end{align}
For the parallel case, we identify $\zeta=\sigma$. The constants $\zs$
and $\zt$ can also be related to this setup: $\zs=-x\cdot \u$ and
$\zt=y\cdot \u$.

\paragraph{Case $a\cdot\u=0$.}
When $a \cdot \u =0$, a different parametrization of the edges is
needed. This is simply done by reversing the role of $z_0$ and $z_1$
\begin{align}
z_0 &= \sigma, \\
z_1 &= B_0 - C_0 \sigma.
\end{align}
With the same procedure as before, we obtain that
in this parametrization $\zcpm = \bpm \left( 1 - \cpm \sigma \right)$
remains the same, but the constants $\bpm$ and $\cpm$ change. Thus,
\begin{align}
B_0 &= \bdh ,	\qquad 	& \bpm &= c + \bdh , \\
C_0 &= 0 ,	\qquad	& \cpm &= \frac{\pm 1}{c + \bdh} ,
\end{align}
and at the starting point of each edge $\sigma = -\bdu$. When the shift
due to smearing is taken into account, $\bpm$ and $\cpm$ are not
altered. In contrast, at each edge, $\sigma$ is shifted to $\sigma \to -
\bdu + T$. For the parallel case, we again identify $\zeta=\sigma$ and
the constants $\zs$ and $\zt$ in this setup are $\zs = x \cdot u$ and
$\zt = - y \cdot u$.

The structure of the edge integrands after each edge is parametrized
with the appropriate $\sigma$-parameter changes as follows (we use $\to$
instead of $=$ below because some terms proportional to $\ln|\sigma|$
are omitted from the result, as explained further on):
\begin{align}
\notag
	P_3 + P_4 \ln|\zcpm|
	&\to P_5 + P_6\ln|\bpm| + P_7\ln|1-\cpm\sigma| \\
	& \quad {}
		+ P_8 \LL(\cpm\sigma) .
\label{eqsigmaform}
\end{align}
The $P_i(\zcpm;z_1)$ coefficients are polynomial in $\zcpm$ and
Laurent polynomial in $z_1$. Their structure is taken from
Eq.~\eqref{eqzcpmint} in the nonparallel case and directly from
Eq.~\eqref{eqcint} for the parallel case. The function $\LL(x)$ is
defined in terms of the dilogarithm~\cite{dilog,maximon}
\begin{equation}
	\LL(x)	= \Re\{\Li_2(x)\}= -\int_0^x\d{t}\,\frac{\ln|1-t|}{t} .
\end{equation}
After the $\sigma$ substitution, the new coefficients are obviously
Laurent polynomials in $\sigma$. In the $a\cdot\u=0$ case, they are just
polynomial, since in that case $z_1$ is constant and hence independent
of $\sigma$. As written, the coefficient $P_8=0$. However, its inclusion
makes the structure of the expression on the right-hand side
of~\eqref{eqsigmaform} stable under $\sigma$ integration, which
generically changes the value of $P_8$. In the nonparallel case,
$\sigma$ integration need only be carried out once. But in the parallel
case, it may need to be carried out repeatedly to generate the
$F^{[k]}(\zeta)$ functions. It then becomes important to recognize the
stability of the given expression structure.

The $\sigma$ integrals can be done using elementary means, with a
partial exception for the $P_7$ and $P_8$ term. Recall that all the
$P_i$ coefficients are rational, with poles only at $\sigma=0$. Thus, also 
the  $P_5$ and $P_6$ terms are rational and hence have rational integrals,
with the possible exception of terms proportional to $\ln|\sigma|$.
They are omitted from the result for the following reason. The
singularity of the integrand at $\sigma=0$ appears because of the
presence of inverse powers of $z_1$ in the summand of
Eq.~\eqref{eqcint}. However, the corresponding original $c$ integral is
regular at $z_1$ and thus all $z_1=0$ (and hence all subsequent
$\sigma=0$) singularities must cancel in the final sum over the $\pm$
and $c=\pm R$ ranges. The same reasoning explains the exclusion of
$b\cdot\u=0$ singularities as discussed in Sec.~\ref{subsecsing}.
The integral of the $P_7$ term has the same structure up to terms
absorbed by $P_5$, with the exception of simple poles like
\begin{equation}
	\int \d\sigma \, \frac{1}{\sigma} \ln|1-\cpm\sigma|
	= - \LL(\cpm\sigma) ,
\end{equation}
which obviously produce terms absorbed by $P_8$. Using integration by
parts and the above identity, the $P_8$ term also produces an integral
of the same form, up to terms absorbed into $P_5$ and $P_7$.

The final result for the integral $I^{mn}_l(R,T;X,Y)$ defined in
Eq.~\eqref{cstint} can be organized as follows. There are two possible
expressions, one for the case when $X$ and $Y$ are not parallel and one
for the case when they are. In either case, the expression has the
structure of the sum over the values $c=\pm R$, over the $\pm$ indices
carried by ($z_{c\pm}$, $B_{c\pm}$ and $C_{c\pm}$), as indicated in
Eq.~\eqref{eqcint}, and over the parallelogram vertices $z_{\mu\nu}$, as
indicated in Eq.~\eqref{eqLsum} (nonparallel case) or
Eq.~\eqref{eqwithf} (parallel case).  The summand has the structure of
the right-hand side of Eq.~\eqref{eqsigmaform}, with the $P_i$
coefficients computed according the procedure discussed above.

\subsubsection{Singularity structure}
\label{subsecsing}
Recall that ultimately we are interested in obtaining an asymptotic
expansion for small $\mu$, with leading behavior of the form $\mu^i
\ln^j \mu$ for some $i$ and $j\ge 0$. For that, we do not need the full
dependence of $I^{mn}_l(R,T;X,Y)$ on $R$ and $T$. We only need the
leading-order expansion for $R,T\to 0$. A priori, it is not completely
obvious what form this expansion will take. However, our explicit
calculations show, based on Eq.~\eqref{eqsigmaform}, that it is possible
to expand in products of powers of $S$, $\ln|S|$ and $\sgn S = S/|S|$,
where $S$ is $R$, or $R\pm T$. Note that the form of such an expansion
is stable under differentiation with respect to $R$ or $T$, provided we
supplement it with terms proportional to $\delta(S)$. Recall that such
differentiations will be necessary in the evaluation of ``part~II'' in
Eq.~\eqref{eqwithunderbrace}, described in step~1 of
Sec.~\ref{secRTsmearing}.  In this section, we describe how these
expansions are carried out and tabulated for later lookup during the
final smearing phase described in Sec.~\ref{secRTsmearing}.

The $R,T$ expansion can be carried out mechanically with computer
algebra using the following simple trick. We replace $R\to \epsilon R$,
$T\to \epsilon T$, where $\epsilon$ is a symbolic parameter and expand
in powers of $\epsilon$ and $\ln\epsilon$. After truncating at the
desired order and setting $\epsilon\to 1$, for each term of the
resulting expression, we use pattern matching to extract its structure
(the $R,T$-independent coefficient, the value of $S$ and the powers in
$S^i \ln^j|S| (\sgn S)^k$). So, the result of each expansion is stored
in structured form. Rational and logarithmic expressions can be
efficiently expanded by \textsc{Mathematica} as they are. But the
dilogarithm $\LL(x)$ poses a few problems because of the need to select
a specific branch at $x=\pm\oo$ and $x=1$.  To circumvent this issue, if
we expect to expand about these arguments, we first use one of the
following identities~\cite{maximon} and exploit the fact that $\LL(x)$
is analytic at $x=0$:
\begin{align} 
\LL(1/x) &= - \LL(x) -\frac{1}{2} \ln^2 \vert x \vert + \frac{\pi^2}{12}
		+ \frac{x}{|x|}\frac{\pi^2}{4} , \label{eqdilog1} \\
\LL(1-x) &= -\LL(x) -\ln \vert 1-x \vert \ln \vert x \vert  + \frac{\pi^2}{6} . \label{eqdilog2}
\end{align}

\begin{figure}[h]
\includegraphics[scale=1]{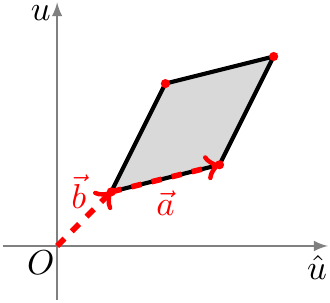}
\caption{Illustration of the role of the vectors $a$ and $b$ defined in
	the text. The vertices of the parallelogram are $z_{\mu\nu}$ and each
	side is a multiple of either $x$ or $y$, defined in
	Eq.~\eqref{eqzijdef}.}
\label{figaandbvector}
\end{figure}

Consider the expression $\ln|A + BS|$. It can clearly have different
leading $S\to 0$ behaviors (or \emph{singularity structure}) depending
on the values of the constants $A$ and $B$. For example, if $A\ne 0$,
then it behaves like $\ln|A| + (B/A)S + \cdots$, while if $A=0$, it
behaves like $\ln|S| + \ln|B|$. The same situation occurs for the
expressions $I^{mn}_l(R,T;X,Y)$ depending on the relative geometry of
the segments $X$ and $Y$. The geometry of these segments is captured by
the geometry of the parallelogram illustrated in
Fig.~\ref{figchangeofvar}. As discussed at the end of the preceding
section (Sec.~\ref{subsecparamseg}), the expression to be expanded
consists of a sum of many terms, each of which depends only on a given
pair of vectors $a$ and $b$ in the $(u,\u)$-plane, where $b$ is a
parallelogram vertex (one of the $z_{\mu\nu}$) and $a$ is one of the
incident parallelogram edges ($\pm x$ or $\pm y$),
cf.~Fig.~\ref{figaandbvector}. The actual dependence appears a
functional dependence on the possible \emph{geometric scalars} generated
from the vectors $a$, $b$, $u$ and $\u$: $a\cdot b$, $a\wedge b$, $a^2$,
$b^2$, $a_\pm$, $b_\pm$, $a\cdot u$, $a\cdot \u$, $\bdu$, $\bdh$. Not
all of these scalars are independent, so for the purposes of some
symbolic manipulations they are expressed in terms of a convenient
independent subset.

\begin{figure*}
\centering
\includegraphics[scale=1]{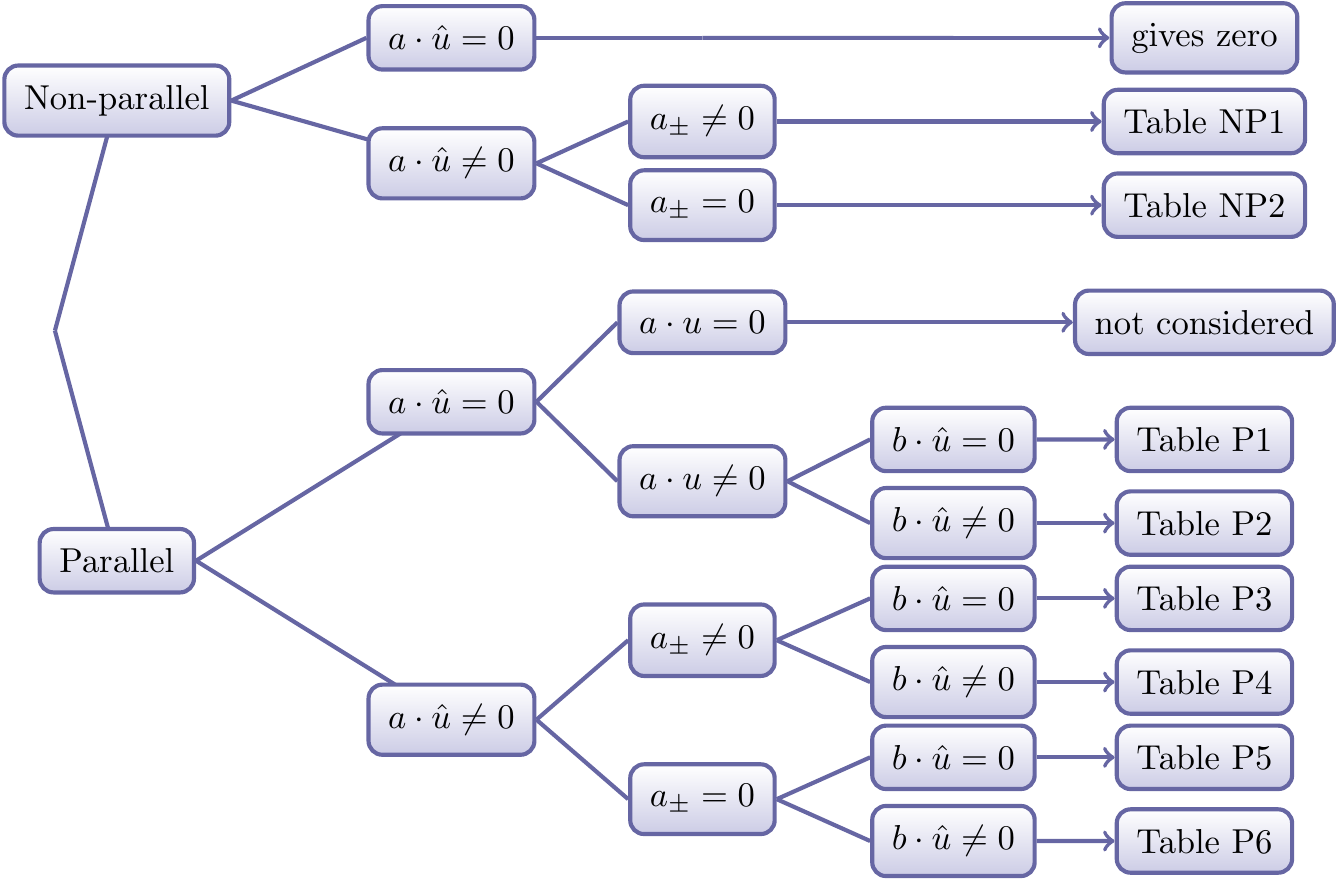}
\caption{Decision tree summarizing the procedure for the nonparallel and parallel cases.}
\label{figdecision}
\end{figure*}

Each end point of the $X$ and $Y$ gives rise to a light cone. Given the
nature of the original integrand (the Hadamard two-point function) in
the definition of $I^{mn}_l(R,T;X,Y)$, it is not surprising that its
singularity structure depends on the position of one segment with
respect to the light cones generated by the other segment or itself. A
detailed study of the expressions in Eq.~\eqref{eqsigmaform} essentially
confirms this expectation. Although, there also appear other
considerations that stem from our particular choices in parametrizing
the edge segment integrals, as described in Sec.~\ref{subsecparamseg}.
The detailed decision trees for determining the singularity structure
for nonparallel and parallel cases are illustrated in
Fig.~\ref{figdecision}. For each possible singularity type, a subset of
the scalars listed in the preceding paragraph is consistently set to
zero, and the $R,T$ expansion is carried out mechanically (as described
before) and the result is stored in structured form in the indicated
table.

After the remaining $R,T$ smearing of ``part~II'', the final answer for
$\expr$ is expected to be of order $1/\mu^2$. We would like to compute a
few subleading terms as well, namely up to and including terms of order
$\O(\mu^0)$. Since the $R,T$ smearing involves applying up to two
derivatives before integrating with respect to the smearing profile, we
must expand in $R$ and $T$ and keep terms up to and including order
$\O(R,T)^2$. However, if we are expanding $I^{mn}_l$ with $l>0$, which
contained $c^l$ in the original integrand, we must keep terms up to and
including order $\O(R,T)^{2+l}$, because the definition of $c$ given in
Sec.~\ref{subsecazimuthal} contains an implicit power of $R$.

As discussed above, the coefficients of the $R,T$ expansion are
functions of various geometric scalars formed from the vectors $a$, $b$,
$u$ and $\u$, and in particular $\bdh$. Some of them contain terms like
$(\bdh)\ln\bdh$, which have well-defined, finite values at $\bdh=0$.
Unfortunately, direct evaluation of such expressions at $\bdh=0$ by
\textsc{Mathematica} produces errors. We have circumvented this problem by
taking the $\bdh\to 0$ limit symbolically beforehand. In the parallel
case, the limit is taken on fully symbolic expressions and is tabulated
separately. However, the same strategy proved to be prohibitively
expensive, with our computational resources, in the nonparallel case,
due to the complexity of the fully symbolic expressions inside the
limit. Instead, we take the limit at a later point of the
calculation, when the numerical values of all the geometric scalars are
available. All of their numerical values are substituted into the
tabulated expression, with the exception of $\bdh$, and the symbolic
limit is taken.

We finish this subsection by briefly summarizing the decision logic
illustrated in Fig.~\ref{figdecision}. We start with an exact formula
for the summand giving $I^{mn}_l(R,T;X,Y)$ for nonparallel or parallel
segments, as in Secs.~\ref{subsecnonprl} and~\ref{subsecprl}. Then, we
check $a\cdot\u = 0$, which decides the edge segment parametrization to
be used, as in Sec.~\ref{subsecparamseg}. In the nonparallel case, the
$a\cdot\u=0$ is trivial, since the integrand is proportional to
$\d{z_1}$, which vanishes in this case. In the parallel case, we further
implicitly assume that $a\cdot u\ne 0$, since otherwise $a=0$, a case
that we do not consider. Next, we check whether $a_\pm \ne 0$ (in our
code labeled `generic') or $a_\pm = 0$ (in our code labeled `special').
Recall that one of $a_+$ or $a_-$ vanishes precisely when $a$ lies on
one or the other branch of the light cone in the $(u,\u)$-plane.
Finally, we check the condition $\bdh=0$. The decision trees in
Fig.~\ref{figdecision} show which table stores the values of the
expansion of $I^{mn}_l(R,T;X,Y)$ with the needed singularity structure.
Each table is indexed by the integers $m,n$ (numbers of iterated segment
integrals) and $l$ (power of $c^l$).

\subsection{Remaining smearing}
\label{secRTsmearing}
Recall the master formula~\eqref{eqwithunderbrace} for
$\tilde{I}^{mn}_K(X,Y)$. As described in the preceding sections, `part~I'
of the calculation has been computed exactly as $I^{mn}_l(R,T;X,Y)$,
expanded for small $R$ and $T$ and an appropriate truncation of the
expansion has been stored in a look-up table. The truncated expansion is
of the form
\begin{equation}
	I^{mn}_l(R,T;X,Y) \sim \sum_f I^{mn}_{l,f}(X,Y) f(R,T) ,
\end{equation}
where each $f(R,T)$ is a product of (possibly singular) powers of $R$,
$R\pm T$, $\ln|R\pm T|$ or $\sgn(R\pm T)$. For simplicity of notation,
we do not show the structure of the truncated expansions in more detail.
The evaluation of `part~II' is carried out algorithmically with the
following steps, which correspond roughly to the summation over the
indices $d$, $\gamma$, $p$ and finally $l$:
\begin{enumerate}
\item
	The summation over $l$ may be carried at any time, so we do it first.
\item
	The $T$ integrals are evaluated by moving all $T$ derivatives from
	$\delta^{(d)}(-T)$ onto the $f(R,T)$ using integration by parts and
	effecting the replacement $T\to 0$. The $\sgn(R\pm T)$ terms generate
	$\delta(R)$'s or derivatives thereof.

	At this point, the summation over $d$ may be carried out.
\item
	Terms proportional to $\delta(R)$ and its derivatives are also evaluated
	using integration by parts and by effecting the replacement $R\to 0$.
	This part of the calculation is then stored separately. It may contain
	terms proportional to $g^{(\gamma)}(0)$.
\item
	In the remaining terms, each $f(R,T)$ has by now been transformed into
	a linear combination of terms of the form $g^{(\gamma)}(R^2) R^i
	\ln^j R$ with powers $i$ such that all integrals are convergent near
	$R=0$. Formal integration by parts (which neglects the boundary terms
	at $R=0$) can bring this expression to the form where each term is now
	$g(R^2) R^i \ln^j R$. However, the powers $i$ may now take values for
	which the integrals diverge near $R=0$. They are to be interpreted as
	distributional integrals, defined by the \emph{Hadamard finite part}
	regularization. 

	At this point, the summation over $\gamma$ may be carried out.
\item
	The distributional $R$ integrals are replaced by moments of the
	smearing function according to the rule
	\begin{equation} \label{eqdefmu}
		\int_0^\infty\d{R} \, g(R^2) \, R^i \ln^j R
		= \mu^{i-2}_{(i,j)} \ln^j \vert \bar{\mu}_{(i,j)} \vert ,
	\end{equation}
	where the numbers $\mu_{(i,j)}$ and $\bar{\mu}_{(i,j)}$ parametrize
	the moments. For simplicity we simply set $\bar{\mu}_{(i,j)} =
	\mu_{(i,j)} = \mu$.

	At this point, the summation over $p$ may be transformed into the
	summation over $i$ in~\eqref{eqwithunderbrace}.
\end{enumerate}

Once the coefficients $P_{\T,d,\gamma,p,l}$ and the truncated expansions
of $I^{mn}_l(R,T;X,Y)$ are known, all of the above operations involve
only elementary algebra on moderate sized expressions and thus can be
efficiently carried out on demand. The result is an expression for
$\tilde{I}^{mn}_K(X,Y)$ in the form given on the last line of
Eq.~\eqref{eqwithunderbrace}. In practice, we truncate the expansions so
that the coefficients $I^{mn}_{\T,i}(\ln\mu;X,Y)$ are known for $i=0,1$
and $2$. A few comments about some of the above steps are in order.

Note that the values $g^{(\gamma)}(0)$, possibly obtained in step 2, can
also be seen as moments of the smearing function, though different from
those defined in Eq.~\eqref{eqdefmu}. In terms of rough scaling, we
expect $g^{(\gamma)}(0) \sim \mu^{-3-2\gamma}$. Thus, the appearance of a
$g(0)$ in the result of the calculation would signify a more singular
leading-order term ($\sim \mu^{-3}$) than is expected by dimensional
analysis and by the form of the last line of~\eqref{eqwithunderbrace}.
Such terms do actually occur in the calculation. Fortunately, and as is
to be expected, they ultimately cancel in the summation over the $X,Y$
segments in Eq.~\eqref{equpdmaster}. This cancellation is taken to be
part of the consistency check on our calculation (Sec.~\ref{secchecks}).

The use of formal integration by parts and the Hadamard finite part
regularization in step 3 are linked. Hadamard finite part (also
\emph{partie finie})
regularization~\cite[Ch.I\textsection{3}]{gelfand-shilov} is defined for
singular integrands $f(R)$ that vanish in the neighborhood of $R=\oo$
and for which there exists a bivariate polynomial $A(x,y)$ such that the
following limit is finite:
\begin{equation}
	P.f.\int_0^\oo f(R)\,\d{R} = \lim_{\epsilon\to 0^+}
		\int_\epsilon^\oo f(R)\, \d{R} - A(\epsilon^{-1},\ln\epsilon) .
\end{equation}
The polynomial $A$ is unique up to the addition of a constant, which may
be absorbed by the replacement $\ln\epsilon\to \ln\epsilon/C$. This
constant may be fixed by requiring that $P.f.\int_0^\oo f'(R)\,\d{R} =
-f(0)$ is always true, provided $f(r)$ vanishes at $R=\oo$. If the
$R$ integrals in ``part~II'' are treated from the start as distributional
integrals~\cite{gelfand-shilov}, with the differentiated smearing functions
$g^{(\gamma)}(R^2)$ playing the role of test functions, then the formal
application of integration by parts produces precisely distributions
regularized according to the Hadamard finite part prescription
	\footnote{An example illustrates this:
\begin{multline*}
\int_0^\infty\!\d{x} \,  g''(x) \ln x =  \lim_{\epsilon\to 0}  \int_\epsilon^\infty\!\d{x} \,  g''(x) \ln x \\
=  \lim_{\epsilon\to 0}   \left( \left[g'(x) \ln x  \right]_\epsilon -
\left[g(x) \frac{1}{x} \right]_\epsilon
-  \int_\epsilon^\infty\!\d{x} \, g(x) \frac{1}{x^2}  \right) \\
= P.f. \int_0^\infty\!\d{x} \, \frac{1}{x^2} g(x) .
\end{multline*}}.
The only addition to formal integration by parts necessary for the
above statement to hold is the rule $1\cdot \frac{\d}{\d{R}} f(R) \to
-\delta(R) f(R)$, rather than $0$. This extra boundary term is then
handled the same as in step 2.

\subsection{Updated master formula for $\expr$}
\label{secupdmaster}
It remains now to evaluate the sums and tensor contractions in the
master formula~\eqref{eqbegincalc} for $\expr$. The tensor contractions
consist of evaluating expressions of the form
\begin{equation}
	r^{Kij}_{nX} \eta_{ij,kl} r^{Lkl}_{mY} \T_{K\cup L} ,
\end{equation}
where $|K|,|L|=0$ or $1$. Reading off the tensorial coefficients from
the explicit expression for $r[h]$, Eq.~\eqref{eqexplicitr}, we can
write them in factored form
\begin{equation} \label{eqrfactor}
	r^{Kij}_{nX}= x^i A^{Kj}_{nX} ,
\end{equation}
where $x$ is the vector corresponding to the segment $X$, with the
orientation indicated by Fig.~\ref{triang-geom-H}. For any tensor basis
element $\T$, we can define the contraction
\begin{equation} \label{eqEdef}
	E_{\T,mn}^{jl}(X,Y)
		= \sum_{K,L} (-)^{|L|} A^{Kj}_{nX} \T_{K\cup L} A^{Ll}_{mX} ,
\end{equation}
with the convention that $\T_J = 0$ for any multi-index $J$ whose size
$|J|$ does not equal the tensor rank of $\T$. We show the structure of
the above multi-index sums explicitly for the needed tensor ranks. Let
$\T^p$ stand for a tensor basis element of rank $p$ (recall also that
$\T^0$ takes only one value, the scalar $1$):
\begin{align}
\label{eqEdef0}
	E_{\T^0,mn}^{jl}(X,Y)
		&= A^j_{mX} A^l_{nY} , \\
\label{eqEdef1}
	E_{\T^1,mn}^{jl}(X,Y)
		&= - A^j_{mX} \T^1_{l_1} A^{l_1\, l}_{nY}
			+ A^{k_1\, j}_{mX} \T^1_{k_1} A^l_{nY} , \\
\label{eqEdef2}
	E_{\T^2,mn}^{jl}(X,Y)
		&= - A^{k_1\, j}_{mX} \T^2_{k_1 l_1} A^{l_1\, l}_{nY} .
\end{align}
The remaining tensor contraction is evaluated using the formula for
$\eta_{ij,kl}$ from
Eq.~\eqref{eqeta}:
\begin{equation} \label{eqetacontract}
	\eta_{ij,kl} \, x^i y^k E^{jl}
	= (x\cdot y) \tr E + E(y,x) - E(x,y) ,
\end{equation}
where $\tr E = \eta_{jl} E^{jl}$ and $E(a,b) = E^{jl} a_j b_l$. The
updated master formula for the quantum variance $\expr$, combining
Eqs.~\eqref{eqbegincalc} and~\eqref{eqwithunderbrace}, can now be
written as follows:
\begin{widetext}
\begin{align}
	\expr
		&= \frac{\ell_p^2}{2\pi}\sum_{KmX}\sum_{LnY}
			(-)^{|L|} r^{Kij}_{mX}\,\eta_{ij,kl}\,r^{Lkl}_{nY} \,
			\frac{2\pi}{\mu^2} \sum_{q=0}^\oo \sum_\T \T_{K\cup L} \, \mu^q \,
				I^{mn}_{\T,q}(\ln\mu;X,Y) , \\
		&= \frac{\ell_p^2}{\mu^2} \sum_{mX} \sum_{nY} \sum_{q=0}^\oo \sum_\T \mu^q \,
			\eta_{ij,kl} \, x^i y^k \, E_{\T,mn}^{jl}(X,Y) \,
			I^{mn}_{\T,q}(\ln\mu;X,Y) \\
\label{equpdmaster}
		&= \frac{\ell_p^2}{\mu^2} \sum_{q=0}^\oo \mu^q
			\sum_{X} \sum_{Y} \eta_{ij,kl} \, x^i y^k
			\sum_{m,n} \sum_\T E_{\T,mn}^{jl}(X,Y) \, I^{mn}_{\T,q}(\ln\mu;X,Y) .
\end{align}
\end{widetext}
Notice, from the above formula, that the final result for $\expr$ may
depend on powers of $\mu$ as well of $\ln\mu$. However, it will be seen
in the next section that $\ln\mu$ \emph{does not} actually appear in the
final result. This fortuitous cancellation can be seen as an explicit
verification of the simple dimensional analysis yielding the $1/\mu^2$
leading singular behavior, as well as a check on the correctness of our
calculations (Sec.~\ref{secchecks}).

This last formula~\eqref{equpdmaster}, directly forms the basis of our
computer algorithm for explicitly evaluating $\expr$ for a fixed
geodesic triangle geometry. We briefly summarize the logic:
\begin{enumerate}
\item
	Load lookup tables for the tensor coefficients $A^{Ki}_{mX}$
	[Eqs.~\eqref{eqexplicitr} and~\eqref{eqrfactor}], tensor basis
	elements $\T$ and polynomial coefficients $P_{\T,d,\gamma,p,l}$
	[Eq.~\eqref{derg} and Table~\ref{tablesmearing}], and truncated
	expansions for $I^{mn}_l(R,T;X,Y)$ [Sec.~\ref{secgeneratingtables},
	Eq.~\eqref{cstint}, Fig.~\ref{figdecision}].
\item
	Construct the segments $X$ of the geodesic triangle geometry as in
	Fig.~\eqref{triang-geom-H}.
\item
	For fixed $m,X,n,Y$ and $\T$, compute the coefficients
	$I^{mn}_{\T,q}(\ln\mu;X,Y)$ [Eq.~\eqref{eqwithunderbrace} and
	Sec.~\ref{secRTsmearing}] and the matrix $E_{\T,mn}^{jl}(X,Y)$
	[Eqs.~\eqref{eqEdef}--\eqref{eqEdef2}].
\item
	Sum over $m$, $n$ and $\T$ in Eq.~\eqref{equpdmaster} and perform the
	remaining tensor contractions using Eq.~\eqref{eqetacontract}.
\item
	Obtain $\expr$ by summing over geodesic triangle geometry segments $X$
	and $Y$ in Eq.~\eqref{equpdmaster} and keeping as many orders in
	$\mu^q$ as available or desired.
\end{enumerate}
The results of explicit computations using the above algorithm are
discussed in the next section.

\section{Results}
\label{secresults}
Here we present the results of our calculation for the leading-order
quantum gravitational corrections to the quantum variance of the
emission time $\tau(s)$ regularized by a finite measurement resolution
scale $\mu$
	\footnote{From which one can easily obtain the quantum corrections to
	the time delay, cf.~Eq.~\eqref{eqemtime}.}.

The experimental geometry is completely determined by two parameters:
the reception time $s$ and the relative velocity $v_{rel}$ between the
worldlines of the lab and the probe, which can also be parametrized by
the (positive) hyperbolic rapidity $\theta$, with $v_{rel}/c=
\tanh(\theta)$. Given these two inputs, in our approximation,
Eqs.~\eqref{eqhsmdef}--\eqref{eqdtau}, the quantum mean and the variance
of quantum fluctuations in the emission time are given by the following
expressions
\begin{align}
\langle \tilde{\tau}(s) \rangle &= \tau_\mathrm{cl}(s) + \O (\ell_p^2) , \\ 
(\Delta\tau)^2 & = \tau_\mathrm{cl}^2 (s) \, \expr +\O (\ell_p^2) , 
\end{align}
where, following Eq.~\eqref{taucl},
\begin{equation}
\tau_\mathrm{cl}(s)= s e^{-\theta} = s \, \sqrt{\frac{1-\frac{v_{rel}}{c}}{1+ \frac{v_{rel}}{c}}}
\end{equation}
and $\langle \tilde{r}^2 \rangle$ is computed by the computer routine as
described in Sec.~\ref{seccalc}.

Using dimensional analysis, as in Sec.~\ref{subsectruncorder}, we can
parametrize the leading contributions to this expectation value as
\begin{equation}
\expr = \frac{\ell_p^2}{\mu^2} \left( \rho_0 + \rho_1 \frac{\mu}{s} + \rho_2\frac{\mu^2}{s^2} + \O\left(\frac{\mu^3}{s^3}\right) \right) + \O\left(\frac{\ell_p^2}{\mu^2}\right), \label{eqr2intermsofrho}
\end{equation} 
where the $\rho_i$ coefficients are in general functions of $v_{rel}/c$.
Note that the result is given to order $\O(\ell_p^2/\mu^2)$ as we did
not include the $\O(h^2)$ term $\tilde{r}_2(h)$ in our calculation of
the variance (see Sec.~\ref{subseclinearized}). The explicit result of
our computer calculation gives
\begin{align}
\label{eqrho0}
	\rho_0 &=
		\frac{1}{v^2}\left(\frac{51}{8}+8v+\frac{141}{8}v^2\right) \\
\notag & \quad {}
		- \frac{1}{v^2}\left(3+4v\right)
				\frac{(1-v^2)}{v}\ln \left(\frac{1+v}{1-v}\right) , \\
\label{eqrho1}
	\rho_1 &= -2 \pi^2 , \\
	\rho_2 &= 0 .
\end{align}
The limiting value at $v=v_{rel}/c<1$ is $\rho_0 = 32$. These
expressions are the main result of our calculation and were, in fact,
the main motivation for it. They deserve a few comments.

It should be mentioned that, in addition to powers of $\mu$ as in
Eq.~\eqref{eqr2intermsofrho}, terms depending on $\ln \mu$ appeared in
intermediate contributions to $\expr$. Remarkably, they all canceled, so
that the final expressions for $\rho_i$ given above depends only on powers of 
$\mu$. The components of the vectors representing the worldline segments
$U$, $V$ and $W$ (Fig.~\ref{triang-geom-H}) are rational functions of
$v_{rel}/c$ and $s$.  Rational expressions in these components appear as
arguments of the graviton Hadamard two-point function and integrals
thereof, as seen Sec.~\ref{seccalc}, which generate further rational and
logarithmic expressions. It is therefore not surprising to see the
$\rho_i$ coefficients of that form, with the $s$ dependence parametrized
away in Eq.~\eqref{eqr2intermsofrho}. However, their simplicity is
striking. Note also that the dependence on $\pi$ is due only to
dilogarithm identities Eqs.~\eqref{eqdilog1}--\eqref{eqdilog2}, since
the overall factor $\frac{1}{2\pi}$ in~\eqref{eqbegincalc} is absorbed
into the normalization factor in the azimuthal angular averaging,
Eqs.~\eqref{eqwavg1} and~\eqref{eqwavg2}.

All the physically relevant information can be glimpsed from the low
velocity approximation for the root-mean-square size of the quantum
fluctuations
\begin{equation}\label{eqDtauapprox}
	\Delta\tau \sim \sqrt{\frac{3}{8}} \left(\frac{c}{v_{rel}}
		\frac{s}{\mu}\right) \ell_p .
\end{equation}
The dimensional scale of the effect is set by the Planck length,
($\ell_p \sim 10^{-35}\, \text{m} \sim 10^{-44} \,\text{s}$). There are
two enhancement factors: the ratio $s/\mu$ of the experimental geometry
and detector resolution scales, and the ratio $c/v_{rel}$ of the speed
of light to the lab-probe relative velocity. We roughly estimated this
enhancement factor in laboratory and cosmological experimental settings
in Table~\ref{tabnumbers}. The large enhancement factors in the
cosmological setting should be taken with a grain of salt. Foremost,
curvature corrections must be added to our Minkowski calculation.
Moreover, in either setting, the divergence of the enhancement factor
for low velocities is rather puzzling, which we discuss next.
\begin{table}[h]
\caption{The enhancement factor $(\frac{c}{v_{rel}}\frac{s}{\mu})$ for
	$\Delta\tau$, Eq.~\ref{eqDtauapprox}, over the Planck scale $\ell_p
	\sim 10^{-44}\,\text{s}$. Detector resolution scale: $\mu =
	10^{-9}\,\text{m}$ (X-ray wavelength). Laboratory scales: $s =
	1\,\text{m}$, $v_{rel} = 1\,\text{m/s}$. Cosmological scales:
	$s=1\,\text{Mpc} \sim 10^{22}\,\text{m}$, $v_{rel}=10^5\,\text{m/s}$
	(Hubble recession velocity at $1\,\text{Mpc}$), $v_{rel} = c/3 \sim
	10^8\,\text{m/s}$ (relativistic velocity).} \label{tabnumbers}
\begin{ruledtabular}
\begin{tabular}{llll}
	& \multicolumn{3}{c}{$v_{rel}$} \\
$s$	& $1 \, \text{m/s}$ & $10^5 \, \text{m/s}$ & $10^8 \, \text{m/s}$  \\ 
\hline 
\rule[0.5ex]{0pt}{2.5ex}%
$1 \, \text{m}$  & $10^{17}$  & $10^{12}$ & $10^{9}$  \\ 
$1 \, \text{Mpc}$  & $10^{39}$ & $10^{34}$ & $10^{31}$  \\  
\end{tabular}
\end{ruledtabular}
\end{table}

\begin{figure}[h]
\includegraphics[scale=.99]{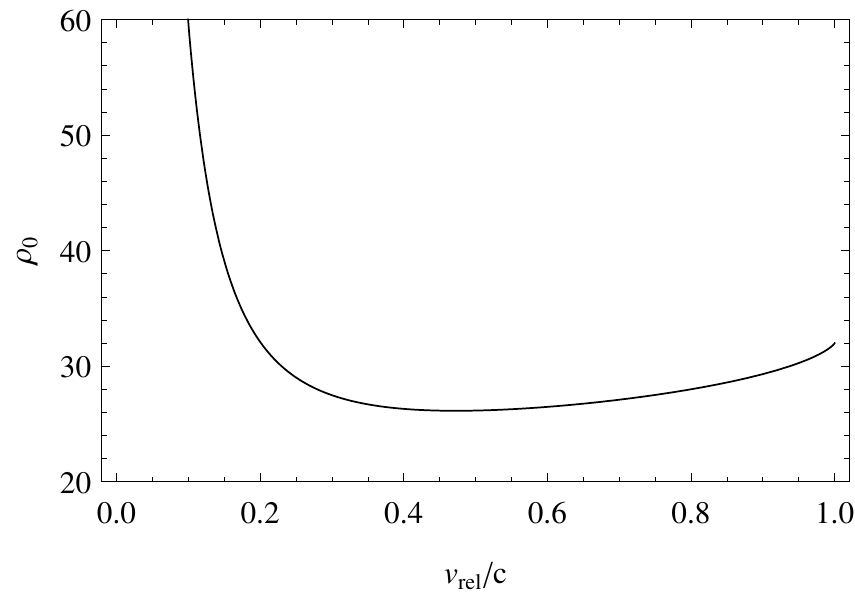}
\caption{Plot of $\rho_0$ versus $v_{rel}/c$ for $\mu=1$. Note the
	divergence at $v_{rel}/c=0$, as discussed in the text.}
\label{fig-plotrho0versusvrel}
\end{figure}

A plot of the coefficient $\rho_0$ versus $v=v_{rel}/c$ is shown in
Fig.~\ref{fig-plotrho0versusvrel}. As is clear from the graph, $\rho_0$
diverges as $1/v^2$ in the limit $v \to 0$. It reaches a minimum around
$v \sim 1/2$ and climbs to the limiting value of $\rho_0 = 32$ as $v \to
1$. The $\rho_0 \sim 1/v^2$ divergence as $v\to 0$ is somewhat puzzling.
The exponent of the divergence can be traced to the value of the
normalization factor $\tau_{\mathrm{cl}}(s)(v\cdot w)$
[Eq.~\eqref{eqclnorm}] that appears in the denominator of the explicit
expression for $r[h]$, Eq.~\eqref{eqr-def}. Classically, the integrals
in the numerator of Eq.~\eqref{eqr-def} all cancel so that $r[h]$
remains finite and in fact goes to $0$ as $v\to 0$. Afterall, there is
no time delay if the lab and probe trajectories coincide. On the other
hand, it seems that the quantum variance of the numerator in
Eq.~\eqref{eqr-def} goes to a nonzero constant as $v\to 0$, thus
resulting in the divergence. While this result is interesting, the
extrapolation of our calculation to $v\to 0$ must be taken with a grain
of salt, since this limit violates our assumption that all sides of the
geodesic triangle must be of size $s$ and that $\mu/s\ll 1$. We cannot
exclude the possibility that the low velocity divergence is naturally
regulated to a finite limit over the range $[0,\mu/s]$ of $v$ in a more
accurate calculation. (The numerical values presented in
Table~\ref{tabnumbers} fall outside these transitional regions as there
the critical velocity is $\mu/s \leq 0.1\,\text{m/s}$.) It is worth
remarking that the dependence of $\rho_0$ on $v_{rel}$ could still be
significantly altered by two factors: a different (hopefully Lorentz
invariant) smearing procedure, and the inclusion of the quadratic
correction $r_2[h]$ to the emission time, both of which would contribute
corrections to the quantum variance at the order $\O(\ell_p^2/\mu^2)$.

Furthermore, a short note on the analytical formula~\eqref{eqrho0}. The
computer calculation was carried out symbolically, but with fixed
(rational) numerical values of $v=v_{rel}/c$ supplied as input. The
analytical expression in terms of $v$ was obtained by a perfect fit to
over 100 data points. 

Finally, a comment on the calculation time. The calculation time can be
divided into two parts (essentially ``part~I'' and ``part~II'' in
Sec.~\ref{seccalc}): generating the tables (which only needs to be done
once) and the explicit calculation of $\expr$ for a given value of $v$.
On a standard computer (AMD 64 Dual Core 2 GHz Processor) the first part
takes approximately 45 minutes (once) and the second part takes about 20
minutes (per value of $v$). The most time consuming part in this latter
calculation is the expansion in $\bdh$ for nonparallel line segments.
If this expansion could also have been tabulated, the calculation time
for $\expr$ would be drastically reduced. 

\subsection{Checks on results}
\label{secchecks}
We implemented several checks to make sure that we can be confident that
the result presented is correct. First of all, the variance of any
physical observable needs to be positive. It is obvious from the graph
in Fig.~\ref{fig-plotrho0versusvrel} that $\expr$ is always positive and
this serves as a first check on our result. In addition, as was remarked
in Appendix~\ref{applinobsv}, parts of the expression of $r$, viz.\ $H$
and $J$ of Eq.~\eqref{eqr-def}, are independently invariant under
linearized diffeomorphism. These parts turn out to satisfy all other
constraints on observables as well and are thus strictly speaking also
observables, although their physical interpretation is not directly
clear. Thus, $H^2$ and $J^2$ and any positive functional thereof should
also be equal to or larger than zero.  This was checked by the same
routine that was used to calculate $\expr$ and indeed it was shown that
$\langle H^2 \rangle\geq0$ and $\langle J^2 \rangle \geq0$.

Second, the results nicely match the predictions made by a simple
dimensional analysis: no terms more divergent than $1/\mu^2$ appear.
In~\cite{khavkine}, it was noted that detailed calculations reveal terms with a more
divergent scaling behavior (these terms are of the form $(\ell_p^2 /
\mu^2) \ln \mu/s$ and $s \ell_p^2 / \mu^3$); however, these terms cancel
in the final result. A generic set of coefficients combining the
$I^{mn}_{\T,i}(\ln\mu;X,Y)$ in a sum over the $X$ and $Y$ segments would
not result in a cancellation of the logarithmic terms. Thus it is unlikely
that the cancellation would happen by accident (i.e.,\ in the case of a
programming error). This observation increases our confidence in the
result, even if the cancellation of the logarithmic terms was not obvious in
advance.

Third, there are two independent parts in our calculation where we
expected intermediate results in the integrals to cancel each other in
the final summations. One of these expected cancellations was (already
mentioned in Sec.~\ref{subseccint} and its reasoning further expanded
upon in Sec.~\ref{subsecparamseg}): the cancellation of terms singular
in $z_1=0$ after all summations are taken into account.  These terms
singular in $z_1$ correspond to poles for $\bdh=0$ after
parametrization. In an independent routine we expanded the results from
the integration in $\bdh$ and checked whether we had rightfully thrown
away all the lower-order terms: the outcome was positive, all lower-order
terms vanish in the final summations over the overall $\pm$-sign,
$c=\pm R$ and the four vertices
	\footnote{It should be noted that our computer did not have enough
	memory to explicitly check one particular case (specifically, $m,n=1$
	and $l=2$). However, there is no reason to expect that the terms in
	this case would not cancel, given the success of other tests.}. %
The other part where we expected cancellations to happen was related to
the integration over $c$. This integration may produce terms that have
the `wrong' powers of $R$, which would eventually lead to terms scaling
as $1/\mu^4$. Fortunately, these terms vanish when the boundaries of the
$c$-summation are taken into account (so the summation over the
$\pm$-sign and $c=\pm R$ is performed). 

Fourth, as a check on the internal consistency of the routine that
handles all integration by parts, the boundary terms produced by
removing all derivatives from the smearing function nicely cancel with
divergent terms in the bulk. This check is discussed in detail at the
end of Sec.~\ref{secRTsmearing}.

Lastly, the integral $\tilde{I}^{mn}_{K}(X,Y)$ has been calculated by
hand for a simple case: $m=n=0$ (hence no iterated integrals) and along
two parallel null segments. This has been done for zero, one
as well as two derivatives on the smearing function, that is, $|K|=0,1,2$.
The results of the calculation by hand match exactly the result produced
by the computer routine and can be found in
Appendix~\ref{apppartialcheck}.

All in all, these partial checks of various aspects of the calculation
make us confident that the result presented is correct. 

\section{Discussion}
\label{secdisc}
In this paper, we have presented the first detailed calculation of the
finite, measurement resolution regulated, quantum fluctuations in a
gauge invariant, nonlocal, operationally defined observable in the
Minkowski linearized quantum gravitational vacuum. As discussed in the
Introduction, this is at least a partial improvement on previous works
in a similar direction~\cite{ford-lightcone, ford-top, ford-focus,
ford-angle,woodard-thesis,tsamis-woodard,ohlmeyer,roura}. Unfortunately, this
calculation is not yet the final word on the matter, due to some
imperfect pragmatic technical choices made along the way. These will be
recalled in more detail below. Still, our calculation can be seen as a
very detailed template for a future, improved calculation or for
generalizations, some examples of which are also discussed below.

The observable we considered is the time delay, induced by metric
fluctuations, between proper time clocks moving at a predetermined
relative velocity and compared using light signals. Generically
speaking, this simple thought experiment setup can be seen as an
idealization of possible laboratory scale, space based, or even
cosmological scenarios. In each of these cases, our results, presented in
Sec.~\ref{secresults}, give an estimate for the size of the quantum
fluctuations and its dependence on the relative velocity of the clocks.
Since this estimate is based only on the rather conservative model of
linearized quantum gravity~\cite{burgess-qg}, it can be used for several
purposes. For instance, the expected size of the fluctuations can be
compared to other sources of noise, like measurement uncertainties and
even intrinsic quantum fluctuations in the measurement equipment, to see
if it is reasonable to expect noticeable quantum gravitational effects in a
given experimental setup. Unfortunately, in the scenarios we have
considered, the quantum gravitational effects seem well below
experimental sensitivity. However, it is not out of the question that
alternative laboratory scenarios~\cite{wbm-coldatom} could bring the size
of such effects closer to the current or future state of the art. Also,
our estimate can be used to contrast predictions of the conservative
linearized gravity model with more exotic ``quantum gravity'' models,
which sometimes (so far unsuccessfully) lay claim to explain anomalies
in cosmological observations, such as the dispersion in the arrival time
of distant $\gamma$-ray burst photons~\cite{grb,ac-grb,hs-pheno}.

Our result for the quantum fluctuation in the time delay shows two odd
features, at least superficially: it is not Lorentz invariant and it
diverges for low relative velocities. It fails to be Lorentz invariant
because it selects a preferred relative velocity (where the fluctuations
are at a minimum) between the two moving clocks (the lab and the probe).
This phenomenon is mostly likely due to an explicitly non-Lorentz
invariant choice of spacetime smearing applied to the graviton field.
This smearing is physically significant, as it is interpreted as the
cumulative effect of the intrinsic (quantum and statistical)
fluctuations in the center of mass coordinates and the limited spacetime
resolution of the measurement equipment. However, to make the
calculation tractable, a pragmatic choice has been made to smear always
in the lab's spatial plane. While this is perfectly acceptable on the
lab worldline, a future follow-up calculation should select a more
realistic smearing profile for the probe and signal worldlines,
preferably in a way that depends only on the local geometry of each
worldline. The divergence in the size of the quantum fluctuations for
small relative velocities, $v_{rel}\to 0$, on the other hand, is more
puzzling. It is not clear what the physical interpretation of this
result would be compatible with the fact that we do not observe such
unbounded fluctuations in the everyday world of slowly relatively moving
macroscopic objects.  There are two chief possibilities that could
explain it as an artifact of our particular calculation. One possibility
is our approximation scheme. It treats the smearing length scale $\mu$
to be much smaller than the length scale $s$ related to the geometric
scale of the experiment. But, as the relative velocity shrinks to zero,
the size $\sim v_{rel} s$ of the signal worldline shrinks as well,
eventually violating the $\mu\ll v_{rel} s$ requirement. Thus, it is
possible that the divergence is resolved into a smooth transition to a
finite limit, which unfortunately cannot be resolved within our
approximation. The other possibility is that the inclusion of a
quadratic correction to the perturbative formula for the time delay
could cancel the low velocity divergence, since that term would
contribute at the same order in $\ell_p/\mu$ as the result computed
here. Its exclusion was again a pragmatic choice made to render the
calculation tractable. A start was made in Appendix~\ref{apppertsol},
where the geodesic and parallel transport equation were calculated to
second order in the gravitational field. These terms are to be included
to get an expression for the time delay that is truly of quadratic
order.  However, presently, the computer routine is not able to handle
some of these terms either because of the appearance of three
derivatives acting on the graviton field (the code handles maximally two)
or integration over the individual geodesic segments is not of a type
that we considered (the $st$ integral is not over the whole unit square,
but over the $0 < s < t < 1$ triangle).  These terms should be fully
taken into account in a follow-up calculation.

Despite the above drawbacks, we believe that the calculation and the
result presented in this work constitute a valuable exercise in the
treatment of phenomenologically meaningful, gauge invariant observables
in quantum gravity. In particular, this calculation and the tools
developed for it can be straightforwardly generalized to handle a large
class of observables that in~\cite{khavkine} were named \emph{quantum astrometric
observables}. This class includes the time delay, angular
blurring~\cite{ford-focus, ford-angle} and other kinds~\cite{roura} of clock
and image distortions induced by the gravitational field in the mutual
observation of a lab and one or more probes. In particular, the details
presented in Sec.~\ref{seccalc} allow an almost immediate generalization
of our triangular setup to more complicated arrangement of lab and probe
worldlines.

Furthermore, note that all intermediate steps of our calculation have
been carried out in position space, rather than momentum space, despite
the translational invariance of the background. The purpose of that
choice was to make a detailed record of the various divergences
encountered in the intermediate steps and their cancellation or
regularization. It is hoped that it can be used to build the intuition
necessary to correctly generalize this kind of calculation to curved
backgrounds. Potential applications of quantum astrometric observables
on curved backgrounds exist in black hole and cosmological scenarios. In
a black hole scenario, one can construct an observable to represent the
size of a black hole and then use it to study the dynamical evaporation
of a black hole with the back reaction on the quantized dynamical
gravitons taken into account. In the cosmological scenario, it would be
a fruitful exercise to explicitly model (some idealization of) the
observations related to the cosmic microwave background (CMB) in a gauge
invariant way.  It is possible that a detailed understanding of the
structure of the corresponding observables may resolve some of the
infrared divergences occurring in graviton-loop corrections to the CMB
power spectrum~\cite{urakawa}.

Finally, as mentioned in the Introduction and discussed in more detail
in~\cite{khavkine}, information about the quantum fluctuations of the time delay
observable in the nonperturbative regime is likely to tell us a lot
about the causal structure of quantum gravity. Unfortunately, our
current perturbative methods obviously do not provide any information in
that regime. It is possible, though, that some exactly solvable or
numerical models with similar phenomenology, like 2+1 dimensional
gravity~\cite{carlip,meusburger} or causal dynamical
triangulations~\cite{loll}, could make nonperturbative calculations
accessible.

\begin{acknowledgments}
I.K.\ would like to thank Renate Loll, Albert Roura, Sabine Hossenfelder
and Paul Reska for their support and helpful discussions and also
acknowledges support from the Natural Science and Engineering Research
Council (NSERC) of Canada and from the Netherlands Organisation for
Scientific Research (NWO) (Project No.\ 680.47.413).
\end{acknowledgments}

\appendix

\section{PERTURBATIVE SOLUTION OF GEODESIC AND PARALLEL TRANSPORT EQUATIONS}
\label{apppertsol}
In this appendix, we summarize some notation needed to define the time
delay observable. We closely follow Sec.~V\,B\,1 and the Appendix
of~\cite{khavkine}, where more details can be found. Though, below, we extend the
solution of the geodesic and parallel transport equations to quadratic
order.

Let $(M,\eta)$ denote the standard four-dimensional Minkowski space and
$x^i$ an inertial coordinate system on it.  The associated standard
tetrad and its dual are $\x_i^a = (\del/\del x^i)^a$ and $\x^i_a=(\d
x^i)_a$, where $a$ is an (abstract) tensor index and $i$ an internal
Lorentz index; $\eta_{ab} \x_i^a \x_j^b = \eta_{ij} =
\mathrm{diag}(-1,1,1,1)_{ij}$. Any other dual pair of orthonormal
tetrads, $e_i^a$ and $e^i_a$, can be specified by applying a local
general linear transformation to the standard ones:
\begin{equation}\label{tetref}
	e_i^a = \bar{T}^{i'}_i \hat{x}^a_{i'},
	~ e^i_a = T^i_{i'} \hat{x}^{i'}_a,
\end{equation}
where $T$ and $\bar{T}$ are spacetime-dependent invertible matrices,
such that $\bar{T}=T^{-1}$. We choose to parametrize them as
$T=\exp(h)$, where $h^i_j$ is an arbitrary matrix. We also write $h_{ij}
= \eta_{ij'} h^{j'}_j$ and call it the \emph{graviton field}. The tetrad
specifies a Lorentzian metric $g_{ab} = \eta_{ij} e^i_a e^j_b$.

Let $O\in M$ be the origin of coordinates and $\hat{e}_i^a$ be a special
tetrad at $O$ (the \emph{lab frame}). Its discrepancy from the tetrad
fields evaluated at $O$ is denoted by
\begin{equation}\label{labref}
	\hat{e}^a_i = (T_O)^{i'}_i \hat{x}^a_{i'} ,
	\quad
	\hat{e}^a_i = L^{i'}_i e^a_{i'} ,
\end{equation}
where $T_O$ could be an arbitrary invertible transformation, but $L$ is
a Lorentz transformation, $L^{i'}_i L^{j'}_{j} \eta_{i'j'} = \eta_{ij}$,
which we parametrize as $L = \exp(h_O)$.

A worldline $\gamma(t)$ is described by its coordinates
$\gamma^i(t)=x^i(\gamma(t))$. Its tangent vector is denoted
$\dot{\gamma}(t)^a$. Knowledge of the tangent vector allows one to
recover the curve as follows
\begin{equation}
	\int_{t_1}^{t_2} \d{t}\, \dot{\gamma}^a(t) \x^i_a
	= \int_{\gamma(t_1)}^{\gamma(t_2)}\d{x}^i
	= \gamma^i(t_2) - \gamma^i(t_1) .
\end{equation}
For convenience, all curves are affinely parametrized from $0$ to $1$.
Thus, the length of a timelike geodesic is equal to the length of its
initial tangent vector.

\begin{figure}
\includegraphics[scale=1]{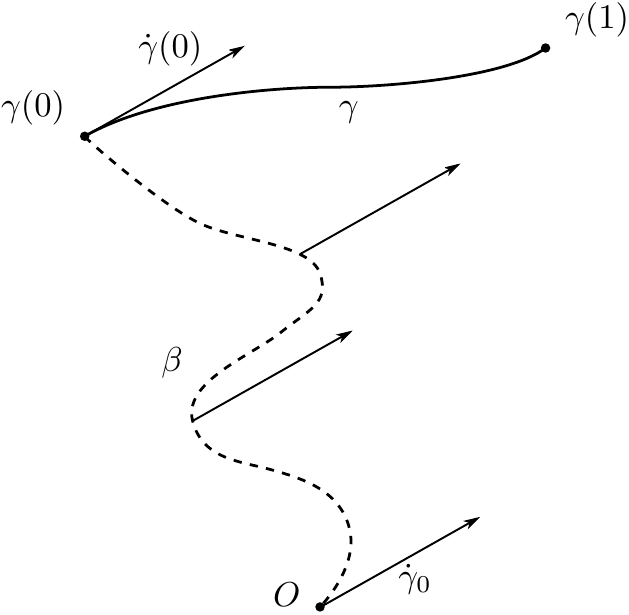}
\caption{%
A geodesic $\gamma$ is defined by its initial point $\gamma(0)$ and
initial tangent vector $\dot{\gamma}(0)$. The initial point itself is
specified as the final point $\gamma(0)=\beta(1)$ of another curve
$\beta$ which starts at the origin. The initial tangent vector can then
be specified by its inverse image $\dot{\gamma}_0\in T_OM$ under
parallel transport along $\beta$.%
}
\label{geodesic-geom}
\end{figure}

A geodesic $\gamma(t)$ is completely specified by its point of origin
$\gamma(0)$ and its initial tangent vector $\dot{\gamma}^a(0)$, while a
$\gamma$-parallel-transported vector $v^a(t)$ is specified by its
initial value $v^a(0)$ at $\gamma(0)$. Again, for convenience in further
calculations, all such initial data are specified with reference to some
given curve $\beta$, with $\beta(0)=O$. Namely, the point of origin is
$\gamma(0)=\beta(1)$, the initial tangent vector $\dot{\gamma}^a(0)$ is
the $\beta$-parallel-transported image of a vector $\dot{\gamma}^a_O =
\dot{\gamma}^i_O \hat{e}^a_i$, and the initial value $v^a(0)$ is the
$\beta$-parallel-transported image of a vector $v^a_O = v^i_O
\hat{e}^a_i$ (cf.~Fig.~\ref{geodesic-geom}).

Let $\gamma(t)$ be a parametrized spacetime curve and $v_\alpha^a(t)$,
$\alpha=0,1,2,3$, an orthonormal tetrad along it. Its components
$v^i_\alpha(t)$ in the basis of the spacetime tetrad are given by
$v^a_\alpha(t) = v^i_\alpha(t) e^a_i(\gamma(t))$. The pair
$(\gamma,v^a_\alpha)$ is a geodesic with a parallel-transported
orthonormal frame on it if it satisfies the following conditions
\begin{align}
	\dot{\gamma}(t)^a &= v^a_0(t), \\
	\dot{\gamma}(t)^a \nabla_a v^c_\alpha(t) &= 0.
\end{align}
The coordinate components of the velocity are $\dot{\gamma}(t)^a \x^i_a
= \dot{\gamma}(t)^a e_a^{j} \bar{T}^i_{j}$. Hence, in coordinate and
tetrad components, the geodesic and parallel transport equations become
\begin{align}
\label{geoeq}
	\dot{\gamma}^i &= v^j_0 \bar{T}^i_j,\\
\label{pteq}
	\dot{v}^k_\alpha &= -v^i_0 \omega\indices{_i^k_j} v^j_\alpha,
\end{align}
where $\eta_{kk'}\omega\indices{_i^{k'}_j} = \omega_{ikj} =
\omega_{i[kj]}$ are the Ricci rotation coefficients (Sec~3.4b
of~\cite{wald}). The Ricci rotation coefficients can be computed in
terms of the transformation matrix $T^i_j$:
\begin{align}
	\omega_{ikj}
	&= -\alpha_{i[kj]} + \alpha_{j(ik)} - \alpha_{k(ij)} , \\
	\alpha_{ikj}
	&= \bar{T}^{i'}_i (\del_{i'} T^{l}_{j'}\eta_{lk}) \bar{T}^{j'}_{j} .
\end{align}
The geodesic~\eqref{geoeq} and parallel transport~\eqref{pteq} equations
can be jointly transformed into a system of integral equations
\begin{align}
	\gamma(t)^i
		&= \gamma(0)^i + \int_{0}^t\d{t'}\, \bar{T}(\gamma(t'))^i_j v_0^j(t'), \\
	v^k_\alpha(t)
	&= T\exp\left[
			-\int_0^t\d{t'}v_0(t')^i\omega(\gamma(t'))\indices{_i^k_j}
		\right] v_\alpha^j(0), \\
\label{eqpp-def}
	&= \exp(p_\gamma(t))^k_j v_\alpha^j(0),
\end{align}
where $T\exp({\cdots})$ denotes the time-ordered exponential and the
\emph{parallel propagator} $\exp(p_\gamma(t))^k_j$ is defined implicitly
by the last equation. For brevity, we also use the notation $p_\gamma =
p_\gamma(1)$. In this form, the solution can be directly expanded to any
desired order in $\O(h)$.

The solution is specified by the following triple of input data: a curve
$\beta(s)$ with $\beta(0) = O$ and $\gamma(0) = \beta(1)$, a frame
$u_\alpha^i(s)$ on $\beta(s)$ with $\dot{\beta}^i = u_0^j
\bar{T}^i_j$, and a Lorentz transformation $L^i_j$ with
$v^j_\alpha(0) = L^j_i u^i_\alpha(1)$. Each of the input data, $\beta$,
$u$, $L$, as well as the resulting $\gamma$ and $v$ can be expanded in
powers of $\O(h)$, with the notation
\begin{equation}
	A
	= \overset{(0)}{A}
	+ \overset{(1)}{A}
	+ \overset{(2)}{A}
	+ \O(h^3) .
\end{equation}
\newcommand{\vo}{\overset{(0)}{v}}%
\newcommand{\vi}{\overset{(1)}{v}}%
\newcommand{\vii}{\overset{(2)}{v}}%
\newcommand{\bo}{\overset{(0)}{\beta}}%
\newcommand{\bi}{\overset{(1)}{\beta}}%
\newcommand{\bii}{\overset{(2)}{\beta}}%
\newcommand{\go}{\overset{(0)}{\gamma}}%
\newcommand{\gi}{\overset{(1)}{\gamma}}%
\newcommand{\gii}{\overset{(2)}{\gamma}}%
\newcommand{\wo}{\overset{(0)}{\omega}}%
\newcommand{\wi}{\overset{(1)}{\omega}}%
\newcommand{\wii}{\overset{(2)}{\omega}}%
\newcommand{\uo}{\overset{(0)}{u}}%
\newcommand{\ui}{\overset{(1)}{u}}%
\newcommand{\uii}{\overset{(2)}{u}}%
\newcommand{\To}{\overset{(0)}{\bar{T}}}%
\newcommand{\Ti}{\overset{(1)}{\bar{T}}}%
\newcommand{\Tii}{\overset{(2)}{\bar{T}}}%
\newcommand{\Lo}{\overset{(0)}{\bar{L}}}%
\newcommand{\LI}{\overset{(1)}{\bar{L}}}%
\newcommand{\Lii}{\overset{(2)}{\bar{L}}}%
First, note the expansions
\begin{align}
	T^i_j
	&= \exp(h)^i_j
	= \delta^i_j + h^i_j + \frac{1}{2} h^i_k h^k_j
		+ \O(h^3) , \\
	\bar{T}^i_j
	&= \exp(-h)^i_j
	= \delta^i_j - h^i_j + \frac{1}{2} h^i_k h^k_j
		+ \O(h^3) , \\
\label{alphaexp}
	\alpha_{ikj}
	&= \del_i h_{kj} \\
\notag & \quad {}
		+ \frac{1}{2}[(\del_i h^m_j) h^l_m \eta_{lk}
			+ h^m_j (\del_i h^l_m) \eta_{lk}] \\
\notag & \quad {}
		- h^{i'}_i (\del_{i'} h^l_j) \eta_{lk}
		- (\del_i h^l_{j'}) \eta_{lk} h^{j'}_j \\
\notag & \quad {}
		+ \gi{}^m \del_m\del_i h^l_j \eta_{lk}
		+ \O(h^3) , \\
	v_\alpha^j(0)
	&= \Lo{}^j_i \uo_\alpha{}^i
		+ \Lo{}^j_i \ui_\alpha{}^i + \LI{}^j_i \uo_\alpha{}^i  \\
\notag & \quad {}
		+ \Lo{}^j_i \uii_\alpha{}^i + \LI{}^j_i \ui_\alpha{}^i
		+ \Lii{}^j_i \uo_\alpha{}^i + \O(h^3) ,
\end{align}
where, in Eq.~\eqref{alphaexp}, $\alpha_{ikj}$ stands for
$\alpha_{ikj}(\gamma(t))$ and all terms on the right-hand side are
evaluated at $t$ or $\go(t)$. For simplicity of notation, we write
\begin{equation}
	A(\go(t)) = A(t) \quad\text{and}\quad
	\overset{(n)}{v}{}_\alpha^i(0) = \overset{(n)}{v}{}_\alpha^i .
\end{equation}
To quadratic order in $\O(h)$, the solutions of the geodesic and
parallel transport equation, Eqs.~\eqref{geoeq} and~\eqref{pteq}, are
given in Eqs.~\eqref{eqv2} and~\eqref{eqgam2} below. 
To keep the structure of the resulting expressions manageable, the terms
are displayed hierarchically. The hierarchy is laid out as follows: (1)
increasing total order in $\O(h)$, (2) decreasing $\O(h)$ order in
inputs ($\beta$, $u$, $L$), (3) increasing number of integrals,  (4)
increasing number of derivatives. The subequations (a), (b) and (c)
refer, respectively, to $\O(h^0)$, $\O(h^1)$ and $\O(h^2)$ terms of the
expansion.

As can be seen from the explicit form of this expansion, there are
several problematic terms appearing at quadratic order that cannot be
accommodated by the evaluation algorithm described in this paper, if
they were to be included as corrections to the quantum variance operator
$r[\tilde{h}]^2$. They are marked by square brackets. These terms are of
the form $\int^t\d{t_1} \int^{t_1}\d{t_2} A[h](t_1) B[h](t_2)$. Our
algorithm would only be able to handle this expression if the upper
limit of the inner integral were also $t$ instead of $t_1$.

\begin{widetext}
\begin{subequations}
\begin{align}
	v_\alpha^k(t)
	&=  \vo{}_\alpha^k \\
\label{eqv2}
       &\quad {}
		+ \vi{}_\alpha^k
		- \int^t\d{t_1}\vo{}_0^i \wi(t_1)\indices{_i^k_j} \vo{}_0^j \\
       &\quad {}
		+ \vii{}_\alpha^k
		- \int^t\d{t_1} \vo{}_0^i \wi(t_1)\indices{_i^k_j} \vi{}_\alpha^j
		+ \left[ \int^t\d{t_1}\int^{t_1}\d{t_2}
			\vo{}_0^{i_1} \wi(t_1)\indices{_{i_1}^k_{j_1}}
			\vo{}_0^{i_2} \wi(t_2)\indices{_{i_2}^{j_1}_{j_2}} \vo{}_\alpha^{j_2} \right ]\\
\notag &\quad {}
		- \int^t\d{t_1} \vi{}_0^i \wi(t_1)\indices{_i^k_j} \vo{}_\alpha^j
		- \int^t\d{t_1} \vo{}_0^i \bi(1)^l
		  (\del_l\wi)(t_1)\indices{_i^k_j} \vo{}_\alpha^j \\
\notag &\quad\quad {}
		- \int^t\d{t_1}\int^{t_1}\d{t_2} \vo{}_0^i \delta^l_{j_2} \vi{}_0^{j_2}
		  (\del_l\wi)(t_1)\indices{_i^k_j} \vo{}_\alpha^j
		- \left[ \int^t\d{t_1}\int^{t_1}\d{t_2} \vo{}_0^i \Ti(t_2)^l_{j_2} \vo{}_0^{j_2}
		  (\del_l\wi)(t_1)\indices{_i^k_j} \vo{}_\alpha^j \right] \\
\notag &\quad\quad\quad {}
		+ \left[ \int^t\d{t_1}\int^{t_1}\d{t_2}\int^{t_2}\d{t_3} \vo{}_0^i \delta^l_{k_3} \vo{}_0^{i_3} \wi(t_3)\indices{_{i_3}^{k_3}_{j_3}} \vo{}_0^{j_3}
		  (\del_l\wi)(t_1)\indices{_i^k_j} \vo{}_\alpha^j \right] \\
\notag &\quad {}
		+ \left[ \int^t\d{t_1}\int^{t_1}\d{t_2}
			\vo{}_0^{i_2} \wi(t_2)\indices{_{i_2}^{i_1}_{j_2}} \vo{}_0^{j_2}
			\wi(t_1)\indices{_{i_1}^k_{j_1}} \vo{}_\alpha^{j_1} \right]
		- \int^t\d{t_1} \vo{}_0^i \wii(t_1)\indices{_i^k_j} \vo{}_\alpha^j \\
\notag &\quad {}
		+ \O(h^3) , 
\end{align}
\end{subequations}
\begin{subequations}
\begin{align}
	\gamma(t)^i
	&= \bo(1)^i + \int^t\d{t_1} \delta^i_j \vo{}_0^j \\
\label{eqgam2}
       &\quad {}
		+ \bi(1)^i + \int^t\d{t_1} \delta^i_j \vi{}_0^j
		+ \int^t\d{t_1} \Ti(t_1)^i_j \vo{}_0^j \\
\notag &\quad\quad {}
		- \int^t\d{t_1}\int^{t_1}\d{t_2} \delta^i_{k_2}
			\vo{}_0^{i_2} \wi(t_2)\indices{_{i_2}^{k_2}_{j_2}} \vo{}_0^{j_2} \\
       &\quad {}
		+ \bii(1)^i + \int^t\d{t_1} \delta^i_j \vii{}_0^j
		- \left[ \int^t\d{t_1}\int^{t_1}\d{t_2} \Ti(t_1)^i_{k_2}
			\vo{}_0^{i_2}\wi(t_2)\indices{_{i_2}^{k_2}_{j_2}} \vo{}_0^{j_2} \right] \\
\notag &\quad\quad\quad {}
		+ \int^t\d{t_1} \bi(1)^l (\del_l\Ti)(t_1)^i_j \vi{}_0^j
		+ \int^t\d{t_1} \Tii(t_1)^i_j \vo{}_0^j \\
\notag &\quad\quad\quad\quad {}
		+ \int^t\d{t_1}\int^{t_1}\d{t_2} \delta^l_{j_2} \vi{}_0^{j_2} (\del_l\Ti)(t_1)^i_j \vi{}_0^j
		+ \left[ \int^t\d{t_1}\int^{t_1}\d{t_2} \Ti(t_2)^l_{j_2} \vo{}_0^{j_2} (\del_l\Ti)(t_1)^i_j \vi{}_0^j \right] \\
\notag &\quad\quad\quad\quad {}
		- \int^t\d{t_1}\int^{t_1}\d{t_2}\int^{t_2}\d{t_3} \delta^l_{k_3} \vo{}_0^{i_3} \wi(t_3)\indices{_{i_3}^{k_3}_{j_3}} \vo{}_0^{j_3} (\del_l\Ti)(t_1)^i_j \vi{}_0^j \\
\notag &\quad\quad {}
		- \int^t\d{t_1}\int^{t_1}\d{t_2} \delta^i_{k_2}
			\vo{}_0^{i_2} \wi(t_2)\indices{_{i_2}^{k_2}_{j_2}} \vi{}_0^{j_2} \\
\notag &\quad\quad\quad {}
		+ \left[ \int^t\d{t_1}\int^{t_1}\d{t_2}\int^{t_2}\d{t_3} \delta^i_{k_2}
			\vo{}_0^{i_2} \wi(t_2)\indices{_{i_2}^{k_2}_{j_2}}
			\vo{}_0^{i_3} \wi(t_3)\indices{_{i_3}^{j_2}_{j_3}} \vo{}_0^{3_2} \right] \\
\notag &\quad\quad {}
		- \int^t\d{t_1}\int^{t_1}\d{t_2} \delta^i_{k_2}
			\vi{}_0^{i_2} \wi(t_2)\indices{_{i_2}^{k_2}_{j_2}} \vo{}_0^{j_2}
		- \int^t\d{t_1}\int^{t_1}\d{t_2} \delta^i_{k_2}
			\vo{}_0^{i_2} \bi(1)^l
				(\del_l\wi)(t_2)\indices{_{i_2}^{k_2}_{j_2}} \vo{}_0^{j_2} \\
\notag &\quad\quad\quad {}
		- \int^t\d{t_1}\int^{t_1}\d{t_2}\int^{t_2}\d{t_3} \delta^i_{k_2}
			\vo{}_0^{i_2} \delta^l_{j_3} \vi{}_0^{j_3}
				(\del_l\wi)(t_2)\indices{_{i_2}^{k_2}_{j_2}} \vo{}_0^{j_2} \\
\notag &\quad\quad\quad {}
		- \left[ \int^t\d{t_1}\int^{t_1}\d{t_2}\int^{t_2}\d{t_3} \delta^i_{k_2}
			\vo{}_0^{i_2} \Ti(t_3)^l_{j_3} \vo{}_0^{j_3}
				(\del_l\wi)(t_2)\indices{_{i_2}^{k_2}_{j_2}} \vo{}_0^{j_2} \right] \\
\notag &\quad\quad\quad {}
		+ \left[ \int^t\d{t_1}\int^{t_1}\d{t_2}\int^{t_1}\d{t_3}\int^{t_3}\d{t_4} \delta^i_{k_2}
			\vo{}_0^{i_2} \delta^l_{k_4} \vo{}_0^{i_4} \wi(t_4)\indices{_{i_4}^{k_4}_{j_4}} \vo{}_0^{j_4}
				(\del_l\wi)(t_2)\indices{_{i_2}^{k_2}_{j_2}} \vo{}_0^{j_2} \right] \\
\notag &\quad\quad {}
		+ \left[ \int^t\d{t_1}\int^{t_1}\d{t_2}\int^{t_2}\d{t_3} \delta^i_{k_2}
			\vo{}_0^{i_3} \wi(t_3)\indices{_{i_3}^{i_2}_{j_3}} \vo{}_0^{j_3}
			\wi(t_2)\indices{_{i_2}^{k_2}_{j_2}} \vo{}_0^{j_2} \right] \\
\notag &\quad\quad\quad {}
		- \int^t\d{t_1}\int^{t_1}\d{t_2} \delta^i_{k_2}
			\vo{}_0^{i_2} \wii(t_2)\indices{_{i_2}^{k_2}_{j_2}} \vo{}_0^{j_2} \\
\notag &\quad {}
		+ \O(h^3) .
\end{align}
\end{subequations}
\end{widetext}

\section{Linearized expression for the time delay observable}
\label{applinobsv}
\begin{figure}
\includegraphics[scale=1]{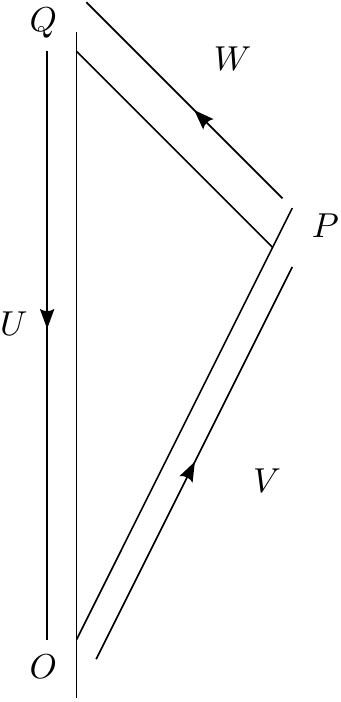}
\caption{%
Illustration of the geometry and orientation of the $U$, $V$ and $W$
segments.
}
\label{triang-geom-H}
\end{figure}

In this appendix, we use the perturbative solution of the geodesic and
parallel transport equations, obtained in Appendix~\ref{apppertsol}, to
find an explicit linearized expression for the time delay observable,
which was defined implicitly in Sec.~\ref{sectimedelay}. We summarize
below the relevant results, whose detailed derivation can be found in
Sec.~V of~\cite{khavkine}.

First, we need to briefly recall some notation introduced in
Appendix~\ref{apppertsol} and introduce some more. Recall that a
lab-equipped spacetime $(M,g,O,\hat{e})$ defines a geodesic triangle
$OPQ$ in $(M,g)$, as illustrated in Fig.~\ref{figexpsetup}. Minkowski
space defines a special lab-equipped spacetime $(M,\eta,0,\x)$, for
which the geodesic triangle and its geometry, including the
corresponding emission and delay times, can be computed explicitly. We
parametrize the deviation of $(M,g,O,\hat{e})$ from $(H,\eta,0,\x)$ with
the space-time dependent general linear transformation $T=\exp(h)$ and
the Lorentz transformation $L=\exp(h_O)$ at $O$, according to
Eqs.~\eqref{tetref} and~\eqref{labref}.

Denote the sides of the geodesic $OPQ$ triangle by the triple
$(\tilde{V},\tilde{W},\tilde{U})$, oriented as shown in
Fig.~\ref{triang-geom-H}. The corresponding initial tangent vectors,
which can be used to define these geodesic segments according to the
scheme of Appendix~\ref{apppertsol}, illustrated in
Fig.~\ref{geodesic-geom}, are $(\tilde{t}\tilde{v}^a, \tilde{w}^a,
-s\tilde{u}^a)$, where $\tilde{w}^a$ is null, $\tilde{u}^a$ is a unit
vector, $\tilde{v}^a = v^i \hat{e}^a_i$ and $\tilde{t} = \tau_v(s)$ is
the emission time, with $s$ and $v^i$ fixed
by the experimental protocol. In Minkowski space, these specialize to
$(V,W,U)$ and $(\tau_\mathrm{cl} v^a, w^a, -su^a)$, where
\begin{align}
	u^a &= u^i\x^a_i = \x^a_0 , \\
	w^a &= s(u^i - e^{-\theta}v^i) \x^a_i , \\
\label{taucl}
	\tau_\mathrm{cl}(s) &= s e^{-\theta}
		= s \sqrt{\frac{1-\frac{v_{rel}}{c}}{1+\frac{v_{rel}}{c}}} ,
\end{align}
with the probe rapidity $\theta$, or equivalently its relative velocity
$v_{rel}$, defined by
\begin{equation}
	u\cdot v = -\cosh\theta = -\frac{1}{\sqrt{1-(\frac{v_{rel}}{c})^2}}.
\end{equation}
We parametrize the analogous non-Minkowski objects as
\begin{align}
	\tilde{t} &= e^{\tilde{r}} \tau_\mathrm{cl}(s) , \label{eqtauinr} \\
	\tilde{v}^a &= v^i \hat{e}^a_i = e^a_i \exp(h_O)^i_j v^j , \\
	\tilde{u}^a &= e^a_i [\exp(p_U)\exp(p_W)\exp(p_V)]^i_j \exp(h_O)^j_k u^k , \\
	\tilde{w}^a &= e^a_i \exp(\tilde{q})^i_j w^j ,
\end{align}
where we have used $\exp(p_\gamma)$ to denote the parallel transport
operator along $\gamma$ as defined in Eq.~\eqref{eqpp-def}, and we
parametrized the changes in $\tilde{t}$ and $\tilde{w}$ due to the
curvature by $\exp(\tilde{r})$, where $\tilde{r}$ is a scalar, and
$\exp{\tilde{q}}$ is a Lorentz transformation. These are determined by
the \emph{triangle closure condition} (the requirement that the
$\tilde{U}$ segment ends in $O$ with tangent vector $-su^a$). Since we
are working at linear order, we only need the leading terms in the
expansion of these unknowns
\begin{equation}
	\tilde{q}^i_j = q^i_j[h] + \O(h^2) ,
	\quad
	\tilde{r} = r[h] + \O(h^2) .
\end{equation}
We have the following linearized expression for the emission time
	\footnote{The previously published expression for $r[h]$, in Eq.~(44)
		of~\cite{khavkine}, had a minus sign missing in front of the term
		proportional to $H$.}
\begin{align}
	\tau(s) &= \tau_\mathrm{cl}(s) [1 + r[h] + \O(h^2)] , \\
\label{eqr-def}
	r[h] &= -\frac{w^i J_i - w^i H_{ij} su^j}{\tau_\mathrm{cl}(s) v\cdot w} ,
\end{align}
where $r[h]$ was obtained from the explicitly expressed triangle closure
condition, using the Eqs.~\eqref{eqv2} and~\eqref{eqgam2} (truncated at
linear order). The normalization factor explicitly evaluates to
\begin{equation}\label{eqclnorm}
	\tau_{\mathrm{cl}}(s) v\cdot w
	= \frac{s^2}{2}(1-e^{-2\theta})
	= s^2 \frac{\frac{v_{rel}}{c}}{(1+\frac{v_{rel}}{c})} .
\end{equation}
The $H$ and $J$ terms are given explicitly by the formulas
\begin{widetext}
\begin{align}
\label{H-expr}
	w^i H_{ij} su^j
	&= \sum_{X=V,W,U} \left(
		s w^i u^j \left[ h_{[ij]} \right]_{x_1}^{x_2}
		+ 2 s w^{[i} u^{j]} x^k \int_X\!\d{t}\, \del_i h_{(kj)} \right), \\
\label{J-expr}
	w^i J_i
	&= \sum_{X=V,W,U} \left(
		- w^i x^j \int_X\d{t} \, h_{(ij)}
		+ 2 w^{[i} x^{j]} x^k \int_X^{(1)}\d{t} \, \del_i h_{(kj)}
		+ \sum_{X>Y=V,W,U} 2 w^{[i} x^{j]} y^k \int_Y\d{t} \, \del_i h_{(kj)}
	\right) ,
\end{align}
\end{widetext}
where $\int^{(n)}_X$ denotes the affinely $[0,1]$-parametrized,
$n$-iterated integral over the segment $X$ with tangent vector $x^a$
(similarly for $Y$ and $y^a$). An ordinary integral is $0$-iterated
$\int^{(0)}\d{t}\, f(t) = \int_0^1 \d{t}\, f(t)$ and $\int^{(1)}\d{t}
f(t) = \int_0^1\d{t}\int_0^t\d{t'}\, f(t')$. The segments are implicitly
ordered $V<W<U$. The first summand term in $H$ depends only on the
antisymmetrization $h_{[ij]}$. Since we shall only use a gauge where
$h_{ij}$ is symmetric, it will can always be neglected in the sequel.

The above linearized expression for $\tau(s)$ is invariant under
linearized gauge transformations (in fact each of the $H$ and $J$ terms
is invariant separately), which has been checked explicitly in the
Appendix of~\cite{khavkine}.

\section{Calculation of the graviton two-point function} 
\label{apptwopoint}
In this appendix, we calculate the Hadamard two-point function, $\langle
\{ \hat{h}_{ij}(x) , \hat{h}_{kl}(y) \} \rangle$, for the linearly
quantized graviton field $\hat{h}_{ij}(x)$, which was defined in
Appendix~\ref{apppertsol}. Obviously, this two-point function depends on
the choice of vacuum state used in the expectation value. In a linear
quantum field theory, the choice of vacuum can be effectively made by
identifying a suitable notion of \emph{positive frequency}
\cite{wald-qft,birrel-davies}. The standard, Poincar\'e invariant Fock
vacuum corresponds to positive frequency with respect to any inertial
time coordinate consistent with the time-orientation of our Minkowski
space $(M,\eta)$. With this choice fixed, it is well known that the
Hadamard two-point function is obtained from the field commutator
$[\hat{h}_{ij}(x),\hat{h}_{kl}(y)]$ by flipping the sign of its negative
frequency Fourier modes. Finally, the field commutator is determined by
proportionality to the classical Poisson bracket, which is fully fixed
by the classical Lagrangian and a choice of gauge fixing. Evidently, the
result depends on the choice of gauge fixing. However, if the Hadamard
two-point function is only used to evaluate expectation values of the form
$\langle \{ O_1[\hat{h}], O_2[\hat{h}] \} \rangle$, where $O_1$ and
$O_2$ are linear gauge invariant observables, these expectation values
will not depend on the choice of gauge, nor even on the addition to the
Hadamard two-point function of anything that is annihilated in the
process. This last observation allows us to choose, in the end, a
particularly simple and symmetric expression for the Hadamard two-point
function. All these steps are performed below.

\subsection{Field commutator}
The field commutator is fixed, according to the usual rules of canonical
quantization, by the formula
\begin{equation}
	[\hat{h}_{ij}(x),\hat{h}_{kl}(y)] = i\hbar\, \Pi(h_{ij}(x),h_{kl}(y)) ,
\end{equation}
where we use $\Pi(-,-)$ to denote the classical Poisson bracket to
distinguish it from the quantum anti-commutator $\{-,-\}$. In a gauge
theory, Poisson brackets are usually defined only on gauge invariant
observables, but are essentially fixed by the Lagrangian density. To
extend Poisson brackets to noninvariant observables, like
the field evaluations $h_{ij}(x)$, we must also specify a gauge fixing.
Below, we use the transverse-traceless-radiation condition~\cite[Sec.4.4b]{wald},
which fully fixes the available gauge freedom.

To determine the Poisson brackets, instead of going through a
complicated 3+1 decomposition and the associated constraint analysis, we
follow the \emph{covariant phase space} formalism~\cite{lee-wald}. The
Lagrangian, together with a choice of Cauchy surface, naturally
determines a 2-form on the space of (off-shell) field configurations.
This 2-form, when restricted to the subspace of solutions (on-shell),
becomes presymplectic and independent of the choice of the Cauchy
surface. Further, restricting to the subspace gauge fixed solution,
which we identify with the \emph{physical phase space}, it becomes
symplectic. We explicitly invert this symplectic form to obtain the
Poisson bivector and hence the Poisson brackets.

\subsubsection{Lagrangian}
Since we are interested in linearized gravity, we start with Minkowski
space $(M,\eta)$ and a global inertial coordinate system $x^\mu$
thereon. For our action, we take the Einstein-Palatini action~\cite{deser}, which in coordinates looks like
\begin{align}
\label{ep-action}
S_{EP} &= \int \d^4{x}\, \mathcal{L}_{EP} = \kappa \int\d^4{x} \, \tilde{g}^{\mu \nu} R_{\mu \nu} , \\
R_{\mu\nu} &= \left( \partial_\lambda \Gamma^\lambda_{\mu \nu} - \partial_\nu \Gamma^\lambda_{\mu \lambda} + \Gamma^\lambda_{\nu \mu} \Gamma^\beta_{\lambda \beta} - \Gamma^\lambda_{\beta \mu} \Gamma^\beta_{\lambda \nu} \right) ,
\end{align}
where $R_{\mu \nu}$ the Ricci tensor, built entirely out of the
Christoffel symbols $\Gamma^\lambda_{\mu\nu}$, and $\tilde{g}^{\mu\nu}$
is the inverse densitized metric, i.e.,\ $g_{\mu\nu}
\tilde{g}^{\nu\lambda} = \sqrt{-g}\delta_\mu^\lambda$, with $g$ the
determinant of $g_{\mu \nu}$.

The independent fields in the Einstein-Palatini action are
$\tilde{g}^{\mu\nu}$ and $\Gamma^\lambda_{\mu\nu}$. The Christoffel
symbols are auxiliary (they can be eliminated algebraically through
their own equations of motion) and their elimination immediately
establishes equivalence with the vacuum Einstein equations. It remains
only to fix the overall constant $\kappa$.

To find this constant, we consider the joint gravity-matter action
$S_{EP}+S_M$, where $S_M$ is the action of a point particle, and impose
on it two conditions: in the nonrelativistic limit (i) $S_M$ has the
standard kinetic term $\int\d{t}\,\frac{mv^2}{2}$ and (ii) the equations
of motion reproduce the standard Poisson equation for the Newtonian
gravitational potential of a particle of mass $m$. It is well
known~\cite[Eq.~8.1]{landlif2} that (i) is satisfied by
\begin{equation}
	S_M = -mc \int\!\d{\tau}\,\sqrt{-\dot{\gamma}^\mu(\tau)\dot{\gamma}^\nu(\tau)g_{\mu \nu}} ,
\end{equation}
where $\gamma^\mu(\tau)$ are the coordinates of the particle's
worldline. Variation of the total action %
	\footnote{The variation is simplified when using the chain rule
	$\delta g_{\mu\nu} = \frac{1}{\sqrt{-g}} (-g_{\mu\mu'}g_{\nu\nu'} +
	\frac{1}{2} g_{\mu\nu} g_{\mu'\nu'}) \delta\tilde{g}^{\mu'\nu'}$.} %
yields the Einstein equations in a form equivalent to
\begin{equation}
	R_{\mu\nu} - \frac{1}{2}g_{\mu\nu}R = \frac{c^3}{2\kappa} T_{\mu\nu} ,
\end{equation}
where $R = g^{\mu\nu} R_{\mu\nu}$ is the Ricci scalar and the
stress-energy tensor of the point particle has the expected form
\begin{equation}
	T_{\mu\nu}(x)= \int\d\sigma \, m c^2 u_\mu u_\nu \, \delta(x,\gamma(\sigma)) ,
\end{equation}
with $\d\sigma = \sqrt{-\dot{\gamma}^\nu \dot{\gamma}^\lambda
g_{\nu\lambda}} \d\tau$, $u^\mu = \dot{\gamma}^\mu /
\sqrt{-\dot{\gamma}^\nu \dot{\gamma}^\lambda g_{\nu\lambda}}$ and
$\delta(x,y)\sqrt{-g} = \delta^4(x-y)$, respectively, the proper time
line element, the unit $4$-velocity and the scalar Dirac distribution.
The correct Newtonian limit is recovered, equivalently, (ii) is
satisfied, if $c^3/2\kappa = 8\pi G$~\cite[\textsection{99}]{landlif2} or
\begin{equation} \label{eqkappa}
	\kappa = \frac{1}{16\pi}\frac{c^3}{G}
		= \frac{1}{16\pi} \frac{\hbar}{\ell_p^2} ,
\end{equation}
where $G$ is Newton's gravitational constant and $\ell_p$ is the Planck
length.

\subsubsection{Gauge fixed symplectic form}
Following~\cite{lee-wald}, we define a 2-form $\Omega$ on the
space of (off-shell) field configurations
\begin{equation}
	\Omega = \int_\Sigma \omega ,
\end{equation}
where $\Sigma$ is a (codimension-1) Cauchy surface and $\omega$ is
itself a 2-form on the space of field configurations, valued in
spacetime 3-forms. We call $\omega$ the presymplectic current density.
When restricted to the subspace of solutions (on-shell), it is
space-time closed, $\d\omega = 0$, as well as variationally closed,
$\delta \omega = 0$, where we have used $\delta$ as the \emph{exterior}
field variational derivative. Hence $\Omega$ is independent of $\Sigma$
and is presymplectic on the space of solutions. When dealing with a
gauge theory, as we are now, its restriction $\bar{\Omega}$ to the
subspace of gauge fixed solutions becomes symplectic. We calculate
$\omega$ from the Lagrangian density $\mathcal{L}$ using the following
steps
\begin{align}
	\delta \mathcal{L} &= EL - \d\theta , \\
	\omega &= \delta \theta ,
\end{align}
where we have again used $\delta$ as the exterior field variational
derivative, $EL$ denotes the term proportional to the Euler-Lagrange
equations and $\d\theta$ is the spacetime exact ``boundary term'' that is
usually discarded while varying the action. We call $\theta$ the
presymplectic potential current density.

Starting with the Einstein-Palatini action~\eqref{ep-action}, we find
\begin{align}
	\theta
	&= \kappa \, (\tilde{g}^{\mu\nu} \delta\Gamma^\alpha_{\mu\nu}
			- \tilde{g}^{\mu\alpha} \Gamma^\nu_{\mu\nu})\, \d^3{x_\alpha} \\
	\omega
	&= \kappa \, (\delta\tilde{g}^{\mu\nu}\wedge \delta\Gamma^\alpha_{\mu\nu}
			- \delta\tilde{g}^{\mu\alpha}\wedge \delta\Gamma^\nu_{\mu\nu})\,
			\d^3{x_\alpha} ,
\end{align}
where $\wedge$ denotes the anti-symmetric product of field variational
forms and in our global inertial coordinates $\d{x^\beta}\wedge
\d^3{x_\alpha} = \delta_\alpha^\beta \d^4{x}$. Letting $\Sigma$ be the
hypersurface $t=x^0 = 0$, we have
\begin{equation}\label{eqomega1}
	\Omega = \kappa \int_{t=0} (\delta\tilde{g}^{\mu\nu} \wedge \delta\Gamma^0_{\mu\nu}
		- \delta\tilde{g}^{\mu 0} \wedge \delta\Gamma^\nu_{\mu\nu}) \,
		\d^3{x_0} .
\end{equation}

At this point, in one step, we restrict to gauge fixed solutions and
expand everything to first perturbative order in the graviton field
$h_{(ij)}$, which was defined in Appendix~\ref{apppertsol}, with the
notation $h_{\mu\nu} = h_{(ij)} (\d{x^i})_\mu (\d{x^j})_\nu$:
\begin{align}
	g_{\mu\nu}
		&= \eta_{\mu\nu} + 2 h_{\mu\nu} + \O(h^2) , \\
	\tilde{g}^{\mu\nu}
		&= \eta^{\mu\nu} - 2 h^{\mu\nu} + \eta^{\mu\nu} h^\alpha_\alpha + \O(h^2) , \\
	\Gamma^\alpha_{\mu\nu}
		&= \eta^{\alpha\beta}(\del_\mu h_{\beta\nu} + \del_\nu h_{\beta\mu}
				- \del_\beta h_{\mu\nu}) + \O(h^2) ,
\end{align}
where indices have been raised using $\eta^{\mu\nu}$. On top of the
equations of motion, to fix the full
available gauge freedom, we impose the transverse-traceless-radiation
condition~\cite[Sec.4.4b]{wald}:
\begin{align}
	\Box \, h_{\mu\nu} &= \del^\lambda \del_\lambda h_{\mu\nu} = 0 , \\
	\del^\mu h_{\mu\nu} &= 0 , \\
	h^\mu_\mu & = 0 , \\
	t_\mu h^{\mu \nu} &= h^{0 \nu} = 0 ,
\end{align}
where $t_\mu = (\d{t})_\mu$.

Making use of the above expansions and gauge fixing conditions, the
form~\eqref{eqomega1} restricts to the symplectic form
\begin{equation}\label{eqomega1a}
	\bar{\Omega} = 2 \int_{t=0}\,
		(\delta h^{\mu\nu} \wedge \delta \dot{h}_{\mu\nu}) \, \d^3{x_0},
\end{equation}
where the dot denotes the $\del_0$ derivative.

The general gauge fixed solution can be explicitly written in Fourier
space, with $x=(t,\vx)$, $k=(\omega,\vk)$ and $\omega_k=|\vk|$, as
\begin{multline}\label{eqgensol}
	h_{\mu \nu}(x)
	= \int\frac{\d^3{\vk}}{(2\pi)^3}\, e^{i\vk\cdot\vx} \\
		{} \times ( P^1_{\mu \nu}(k)
			[ \alpha^+_1(k) e^{-i\omega_k t} + \alpha^-_1(k) e^{i\omega_k t} ] \\
		{} + P^2_{\mu \nu}(k)
			[ \alpha^+_2(k) e^{-i\omega_k t} + \alpha^-_2(k) e^{i\omega_k t} ] ) ,
\end{multline}
where $\alpha_i^\pm(k)$ are arbitrary $k$-dependent coefficients and
$P^i_{\mu \nu}$ are \emph{polarization factors} that need to satisfy 
\begin{equation} \label{eqrestr}
\begin{aligned}
	\eta^{\mu \nu} P^i_{\mu \nu} &= 0, \\
	t^\mu P^i_{\mu \nu} &= P^i_{0 \nu} = 0, \\
	k^\mu  P^i_{\mu \nu} &= 0, \\
	P^i_{\mu \nu} P^{j\,\mu \nu} &= \delta^{ij}, \\
	P^1_{\mu \nu}(-k) =  P^1_{\mu \nu}(k), & \quad
	P^2_{\mu \nu}(-k) = -P^2_{\mu \nu}(k),
\end{aligned}
\end{equation}
for $i=1,2$. The first three conditions are directly related to the
gauge fixing, the orthogonality condition ensures that the coefficients
$\alpha^\pm_i$ describe independent polarizations for different $i$, and
the last conditions prescribe their behavior under parity
transformations %
	\footnote{These prescriptions are satisfied, for instance, if $P^1_{\mu \nu} \sim \theta_\mu\theta_\nu + \phi_\mu\phi_\nu$ and $P^2_{\mu \nu} \sim \theta_\mu\phi_\nu + \phi_\mu \theta_\nu$ in polar coordinates.}. %
In the final expression for the Poisson bracket, only the
\emph{projected identity} tensor $P^1_{\mu \nu}P^1_{\lambda \kappa} +
P^2_{\mu \nu}P^2_{\lambda \kappa}$ appears explicitly. So, instead of
finding explicit expressions for the polarization factors and computing
the projected identity from its definition, we simplify the calculation
by expressing it in the most general basis and then restricting the
coefficients using all of the above conditions. As a basis, we use all
rank-4 tensors that could be constructed from $\eta_{\mu \nu}$, $t_\mu$,
and $k_\mu$ that are symmetric under the index exchanges $(\mu \nu)
\leftrightarrow (\kappa \lambda)$, $\mu \leftrightarrow \nu$ and $\kappa
\leftrightarrow \lambda$: 
\begin{widetext}
\begin{align}
P^1_{\mu\nu} P^1_{\kappa\lambda} + P^2_{\mu\nu} P^2_{\kappa\lambda}
&= \frac{1}{2} \left(\eta_{\mu\kappa} \eta_{\nu\lambda} +\eta_{\mu\lambda} \eta_{\nu\kappa} \right) + A \, \eta_{\mu \nu} \eta_{\kappa\lambda} +
B \left(\eta_{\mu \nu} k_\kappa k_\lambda + \eta_{\kappa\lambda} k_\mu k_\nu \right) \\ \notag &\quad {} + C \left(\eta_{\mu \kappa} k_\nu k_\lambda + \eta_{\nu \kappa} k_\mu k_\lambda + \eta_{\mu \lambda} k_\nu k_\kappa + \eta_{\nu \lambda} k_\mu k_\kappa \right)
 	+ D \left(\eta_{\mu \nu} t_\kappa t_\lambda + \eta_{\kappa\lambda} t_\mu t_\nu \right) \\ \notag &\quad {} +  E \left(\eta_{\mu \kappa} t_\nu t_\lambda + \eta_{\nu \kappa} t_\mu t_\lambda + \eta_{\mu \lambda} t_\nu t_\kappa + \eta_{\nu \lambda} t_\mu t_\kappa \right) + 
F \left(\eta_{\mu \nu} k_\kappa t_\lambda + \eta_{\mu \nu} k_\lambda t_\kappa + \eta_{\kappa\lambda} k_\mu t_\nu
  + \eta_{\kappa\lambda} k_\nu t_\mu \right) \\ \notag &\quad {} + G \left(\eta_{\mu\kappa} k_\nu t_\lambda + \eta_{\nu \kappa} k_\mu t_\lambda + \eta_{\mu\lambda} k_\nu t_\kappa +\eta_{\nu \lambda} k_\mu t_\kappa \right)  + H \left(\eta_{\mu\kappa} k_\lambda t_\nu + \eta_{\nu\kappa} k_\lambda t_\mu +  \eta_{\mu\lambda} t_\nu k_\kappa + \eta_{\nu\lambda} t_\mu k_\kappa \right) \\ \notag
&\quad {} + I \, k_\mu k_\nu k_\kappa k_\lambda + J \left(k_\mu k_\nu k_\kappa t_\lambda + k_\mu k_\nu k_\lambda t_\kappa + k_\mu k_\kappa k_\lambda t_\nu + k_\nu k_\kappa k_\lambda t_\mu \right)  \\ \notag &\quad {} + K \left(k_\mu k_\nu t_\kappa t_\lambda + k_\kappa k_\lambda t_\mu t_\nu \right)
  + L \left(k_\mu k_\kappa t_\nu t_\lambda + k_\mu k_\lambda t_\nu t_\kappa + k_\nu k_\kappa t_\mu t_\lambda + k_\nu k_\lambda t_\mu t_\kappa \right) \\ \notag &\quad {} +  M \left(k_\mu t_\nu t_\kappa t_\lambda + k_\nu t_\mu t\kappa t_\lambda + k_\kappa t_\mu t_\nu t_\lambda + k_\lambda t_\mu t_\nu t_\kappa \right)
  + N \, t_\mu t_\nu t_\kappa t_\lambda
\end{align}
\end{widetext}
where the capital letters are constants that will be fixed by the
restrictions in~\eqref{eqrestr}. We have 14 constants: the trace
condition gives four independent constraints and projection onto $t^\mu$
and $k^\mu$ each give ten constraints amounting to a total of 24
constraints. Fortunately, some constraints are redundant and the system
is exactly solvable. Having set $t^2=-1$, we obtain the following
expressions for the constants:
\begin{align*}
A &= -\frac{1}{2}, 									& H&=\frac{-k\cdot t}{2 \left(k^2+ (k \cdot t)^2\right)}, \\
B &= \frac{1}{2 \left(k^2+ (k \cdot t)^2\right)}, 	& I&=\frac{1}{2 \left(k^2+ (k \cdot t)^2\right)^2}, \\
C &= -\frac{1}{2 \left(k^2+ (k \cdot t)^2\right)}, 	& J&=\frac{k\cdot t}{2 \left(k^2+ (k \cdot t)^2\right)^2}, \\
D &= -\frac{k^2}{2 \left(k^2+(k \cdot t)^2 \right)}, & K&=\frac{k^2 + 2(k \cdot t)^2 }{2 \left(k^2+ (k \cdot t)^2\right)^2},\\
E &= \frac{k^2}{2 \left(k^2+(k \cdot t)^2 \right)}, 	& L&=-\frac{k^2}{2 \left(k^2+ (k \cdot t)^2\right)^2}, \\
F &= \frac{k\cdot t}{2\left(k^2+ (k \cdot t)^2\right)}, & M&=-\frac{k^2 (k \cdot t)}{2 \left(k^2+ (k \cdot t)^2\right)^2},\\
G &= -\frac{k\cdot t}{2\left(k^2+ (k \cdot t)^2\right)}, & N&=\frac{k^4}{2 \left(k^2+ (k \cdot t)^2\right)^2}.
\end{align*}
The resulting projected identity is rather long and complicated.
Conveniently, there are some simplifications that can be made. Since
$h_{\mu\nu}$ has gauge degrees of freedom of the form $k_{(\mu}
\zeta_{\nu)}$, all terms that have a similar form can consequently be
gauged away when calculating observables. Additionally, the projected
identity will appear in the graviton two-point function within an
integral over $k$ together with $\delta^{(4)}(k^2)$; hence all terms
that are proportional to $k^2$ will vanish. Ergo, the only nonvanishing
constant is $A=-\frac{1}{2}$ and, for the purpose of calculating with
gauge invariant observables, the projected identity can be taken to be
simply
\begin{align}
P^1_{\mu \nu} P^1_{\kappa \lambda}+ P^2_{\mu \nu} P^2_{\kappa \lambda} &= \frac{1}{2} \left( \eta_{\mu \kappa} \eta_{\nu \lambda} + \eta_{\mu \lambda} \eta_{\nu \kappa} - \eta_{\mu \nu} \eta_{\kappa\lambda} \right) \notag \\
\label{eqprojid}
&\equiv \frac{1}{2} \eta_{\mu \nu,\kappa \lambda}.
\end{align}
Inserting the general gauge fixed solution~\eqref{eqgensol} into the
symplectic form, expression~\eqref{eqomega1a}, gives
\begin{align}
\notag
\bar{\Omega} &= 2\kappa \int_{t=0}\d^3{\vx} \int\frac{\d^3{\vk}}{(2\pi)^3}  \int\frac{\d^3{\vk'}}{(2\pi)^3} \, i\omega_{k'} e^{i(\vk+\vk')\cdot\vx}  \\
\notag
& {} \times \left( P^{1 \,\mu \nu}(k) [ \delta\alpha^+_1(k) e^{-i\omega_k t} + \delta\alpha^-_1(k) e^{i\omega_k t} ] \right. \\
& \quad \left. {} + P^{2 \,\mu \nu}(k) [ \delta\alpha^+_2(k) e^{-i\omega_k t} + \delta\alpha^-_2(k) e^{i\omega_k t} ] \right) \\
\notag
& \wedge \left( P^1_{\mu \nu}(k') [ -\delta\alpha^+_1(k') e^{-i\omega_{k'} t} + \delta\alpha^-_1(k') e^{i\omega_{k'} t} ] \right. \\
\notag
& \quad \left. {} + P^2_{\mu \nu}(k') [ -\delta\alpha^+_2(k') e^{-i\omega_{k'} t} + \delta\alpha^-_2(k') e^{i\omega_{k'}t} ] \right) \\
\notag
&= 4\kappa \int_{\mathbb{R}^3_+}\frac{\d^3{\vk}}{(2\pi)^3} \, i\omega_k \\
\label{eqksymp}
& \qquad {} \times \left( - \delta\alpha_1^+ \wedge \delta \alpha_1^{+*} + \delta\alpha_1^- \wedge \delta \alpha_1^{-*} \right. \\
\notag
& \qquad\quad \left. {} - \delta\alpha_2^+ \wedge \delta \alpha_2^{+*} + \delta\alpha_2^- \wedge \delta \alpha_2^{-*} \right) .
\end{align}
This result requires some explanation. The $\vx$-integration results in
a factor of $\delta^3(\vk+\vk')$, which eliminates the $\vk'$-integral.
Further, the reality of the graviton field, $h^*_{\mu\nu}(x) =
h_{\mu\nu}(x)$, and the parity properties of $P^i_{\mu\nu}(k)$ given
in~\eqref{eqrestr} translate to the following parity properties of the
$\alpha^\pm_i$ coefficients:
\begin{equation} \label{eqalphapar}
\begin{aligned}
\alpha_1^+(k) &= \alpha_1^{-*}(-k), \\
\alpha_1^-(k) &= \alpha_1^{+*}(-k), \\
\alpha_2^+(k) &= -\alpha_2^{-*}(-k), \\
\alpha_2^-(k) &= -\alpha_2^{+*}(-k). 
\end{aligned}
\end{equation}
Taking these parity properties into account, the integration domain can
then be shrunk from all of $\mathbb{R}^3$ to $\mathbb{R}^3_+$, the
half-space satisfying $k_z \ge 0$. Effecting the remaining algebraic
simplifications gives the expression~\eqref{eqksymp}, where the argument of
each $\alpha$-coefficient is $+k$ and hence has been omitted for
conciseness. Each of the $\alpha$-coefficients appearing
in~\eqref{eqksymp} is now independent of the others, at fixed $\vk$ and
at other values of $\vk\in \mathbb{R}^3_+$.

\subsubsection{Gauge fixed Poisson brackets}
If we consider the $\alpha^\pm_i(k)$-coefficients as a complete set of
independent coordinates on the physical phase space of linearized
gravity, then the expression~\eqref{eqksymp} for the symplectic form
shows that they are also canonical. Therefore, it is straightforward to
write down the Poisson bivector $\Pi = \bar{\Omega}^{-1}$:
\begin{align}
\notag
\Pi &= \frac{1}{4\kappa} \int_{\mathbb{R}^3_+}\d^3{k} \, \frac{(2\pi)^3}{ik} \\
\label{eqkpois}
& \qquad {} \times \left( - \del_{\alpha_1^+} \wedge \del_{\alpha_1^{+*}} + \del_{\alpha_1^-} \wedge \del_{\alpha_1^{-*}} \right.\\
\notag
& \qquad\quad \left. {} - \del_{\alpha_2^+} \wedge \del_{\alpha_2^{+*}} + \del_{\alpha_2^-} \wedge \del_{\alpha_2^{-*}} \right) ,
\end{align}
where the field variational vector fields $\del_{\alpha^{\pm}_i}$ for
are dual to the field variational $1$-forms $\delta\alpha^{\pm}_i$.
These vector fields, through the standard action of vectors and
bivectors on functions, also satisfy the following identities, which
follow from the parity properties~\eqref{eqalphapar}:
\begin{align*}
\del_{\alpha_1^+ (k)} \alpha^+_1 (q) &= \delta(\vk-\vq),\\
\del_{\alpha_1^+ (k)} \wedge \del_{\alpha_1^+ (k)^*} \left(  \alpha^+_1 (p),  \alpha^-_1 (q) \right) &= \delta(\vk-\vp) \delta(\vk+\vq), \\
\del_{\alpha_1^+ (k)} \wedge \del_{\alpha_1^+ (k)^*} \left(  \alpha^+_1 (p),  \alpha^+_1 (q) \right) &= 0, \\
\del_{\alpha_1^- (k)} \wedge \del_{\alpha_1^- (k)^*} \left(  \alpha^+_1 (p),  \alpha^-_1 (q) \right) &= - \delta(\vk-\vp) \delta(\vk+\vq), \\
\del_{\alpha_2^+ (k)} \wedge \del_{\alpha_2^+ (k)^*} \left(  \alpha^+_2 (p),  \alpha^-_2 (q) \right) &= - \delta(\vk-\vp) \delta(\vk+\vq), \\
\del_{\alpha_2^- (k)} \wedge \del_{\alpha_2^- (k)^*} \left(  \alpha^+_2 (p),  \alpha^-_2 (q) \right) &= \delta(\vk-\vp) \delta(\vk+\vq).
\end{align*}
Finally, using the above identities, together with the explicit
parametrization~\eqref{eqgensol} of
gauge fixed solutions, the explicit formula~\eqref{eqkpois} for the Poisson
bivector and the simplified expression~\eqref{eqprojid} for the
projected identity tensor, we obtain the Poisson brackets of two
graviton field evaluations
\begin{align}
&\Pi(h_{\mu\nu}(x),h_{\kappa\lambda}(y)) \notag\\
&= \frac{2\pi}{4\kappa i} \int\!\!\frac{\d^4{k}}{(2\pi)^4} \, \delta(k^2) \left( P^1_{\mu \nu} P^1_{\kappa \lambda}+ P^2_{\mu \nu} P^2_{\kappa \lambda} \right) e^{ik\cdot(x-y)} \sgn(\omega) \notag \\
&=\frac{1}{2} \, \frac{2\pi}{4\kappa i} \, \eta_{\mu \nu, \kappa \lambda} \int\frac{\d^4{k}}{(2\pi)^4} \, \delta(k^2) e^{ik\cdot(x-y)} \sgn(\omega), \label{eqhh}
\end{align}
where $\sgn(x)$ is the sign-function and as before $k=(\omega,\vk)$,
with the extra integration over $\omega$ compensated by the
$\delta(k^2)$ factor and the various $e^{\pm i\omega_k t}$ factors
absorbed into the single remaining exponential.

\subsection{Sign flip}
Having computed the Poisson brackets of field evaluations in the gauge
fixed, linear, classical graviton field theory, canonical quantization
uniquely fixes the commutator of the corresponding quantum field
operators:
\begin{equation}
	[\hat{h}_{\mu\nu}(x),\hat{h}_{\kappa\lambda}(y)]
	= i\hbar \Pi(h_{\mu\nu}(x),h_{\kappa\lambda}(y)) .
\end{equation}
As mentioned earlier, it is well known~\cite{wald-qft,birrel-davies} that
the above field commutator is related to the Hadamard two-point function
by a flip of the sign of its negative frequency components:
\begin{align}
\notag
	\MoveEqLeft \langle \{\hat{h}_{\mu\nu}(x),\hat{h}_{\kappa\lambda}(y)\} \rangle \\
	&= i\hbar \, \sgn(i\del_t) \, \Pi(h_{\mu\nu}(x),h_{\kappa\lambda}(y)) \\
\label{eqgrav2ptmomentum}
	&= \frac{\hbar}{2} \, \frac{2\pi}{4\kappa} \,
		\eta_{\mu \nu, \kappa \lambda}
		\int\frac{\d^4{k}}{(2\pi)^4} \, \delta(k^2) e^{ik\cdot(x-y)} \\
\label{eqgravitons2pointfunction}
	&= \eta_{\mu\nu,\kappa\lambda} \, \frac{\ell_p^2}{\pi}
		P \left[ \frac{1}{(x-y)^2} \right] ,
\end{align}
where the symbol $P$ denotes the \emph{Cauchy principal value}
prescription and we have used the identity~\cite{gelfand-shilov}
\begin{equation} \label{eqdistrid}
	\int\frac{\d^4 k}{(2\pi)^4} \, \delta(k^2) e^{ik\cdot x}
	= \frac{2}{(2\pi)^3} P \frac{1}{x^2}
\end{equation}
and the value $\kappa = \frac{1}{16\pi}\frac{\hbar}{\ell_p^2}$ from
Eq.~\eqref{eqkappa}. We should note that Eq.~\eqref{eqdistrid} is an
identity involving two distributions, a Dirac-delta and a Cauchy
principal value, which are strictly defined only when their arguments
have simple zeros and poles, respectively. Unfortunately, this condition
fails precisely at $k=0$ and $x=0$, respectively, so the distributions
are only defined in the complements of these points. However, in
four dimensions, each distribution can be uniquely extended to the
excluded point provided that it remains homogeneous~\cite{gelfand-shilov}.

\section{Calculation of partial check} 
\label{apppartialcheck}
As a partial check on our computer routine, we calculated---for a very
specific case---the smeared integral $\tilde{I}^{00}$ ($|K|=0$)
by hand. In particular, we considered the integral along two coinciding
(hence parallel) 0-iterated null line segments. This case was chosen
because of its fairly simple calculation and limited number of
intermediate steps.  Indeed all results in this appendix are reproduced
by our computer routine.

We calculated the following integral along two parallel null line segments
\begin{equation}
	\tilde{I}^{00} = \int\!\d^4{z} \int_0^1\!\d{s} \int_0^1\!\d{t} \, P \frac{g(z_\perp^2) \delta(u\cdot z)}{\left(x(s) - y(t) - z \right)^2} .
\end{equation}
We consider the null segments $x=y=\lambda(u-\u)$, for some scalar
$\lambda$, so that $x(s)=s x$ with $\lambda$ and $y(t) =ty = tx$. Since
$x^2=y^2=0$ and $z \cdot x = -\lambda (c-T)$, we can rewrite the
denominator as
\begin{align}
	((s-t)x -z)^2 &= z^2 + 2(t-s) x\cdot z \\
		&= R^2 - T^2 - 2\lambda (t-s) (c-T).
\end{align}
Parametrizing the integral in a similar manner as in Sec.~\ref{seccalc}
and rearranging the denominator gives
\begin{multline}
	\tilde{I}^{00} =  - \int_0^{2\pi}\!\d{\phi}
		\underbrace{\int_{-\infty}^{\infty}\!\d{T} \delta^{(d)}(-T) \int_0^\infty\!\d{R} \, R \, g(R^2) \,}_{\text{part II}} \\
	{}\times
		\underbrace{\int^1_0\!\d{s} \int^1_0\!\d{t} \int_{-R}^{R}\!\d{c} P \, \frac{1}{R^2 - T^2 -2\lambda(t-s)(c-T)}}_{\text{part I}}.
\end{multline}
To compute this integral, we follow the same logic that the computer
algebra uses: ``part~I'' first, which amounts to performing the integration
over $c$, $s$ and $t$, and subsequently, ``part~II'', which entails
integration over $R$ and  $T$. The $\phi$ integral merely contributes an
overall factor of $2\pi$. Integration over $c$ yields
\begin{multline}
\int_{-R}^R\!\d{c} \, \frac{1}{R^2 - T^2 - 2\lambda(t-s) (c-T)} \\ = \left[ - \frac{\left( \ln |c-T| + \ln |c+T-2\lambda(t-s)| \right)}{2\lambda(t-s)} \right]_{-R}^R \label{eqappenc1}.
\end{multline}
For simplicity, below, we work with the expression inside the square
brackets and plug in the boundary values $c=\pm R$ at the end. Next,
integration along $s$ and $t$ is performed. The first term
in~\eqref{eqappenc1} gives
\begin{multline*}
\int\!\d{s} \int\!\d{t} \left(- \frac{\ln |c-T|}{2\lambda (t-s)}\right) \\ = -\frac{1}{2\lambda} \ln|c-T| \left[\left[ -(t-s) \left( \ln|t-s| -1 \right) \right] \right],
\end{multline*}
where the square brackets indicate that the boundaries of $s$ and $t$
still need to be inserted. These boundaries correspond to the four
vertices in Fig.~\ref{figchangeofvar}.  The second term
in~\eqref{eqappenc1} yields
\begin{multline*}
\int\!\d{s}\!\int\!\d{t} \left(- \frac{\ln |c+T-2\lambda(t-s)|}{2\lambda (t-s)}\right) \\ = -\frac{\ln |c+T|}{2\lambda} \left[\left[ -(t-s) \left( \ln|t-s| - 1 \right) \right] \right] + \frac{1}{2\lambda} [[(t-s)]] \\
-\frac{1}{2\lambda} \left[\left[ \left( t-s - \frac{c+T}{2\lambda} \right) \ln \Big\vert 1- \frac{2\lambda (t-s)}{c+T} \Big\vert \right. \right. \\
\left. \left. + (t-s) L \left(\frac{2\lambda (t-s)}{c+T} \right) \right] \right].
\end{multline*}
After combining the two equations again and extracting the logarithmic
divergences from the dilogarithm by applying Eq.~\eqref{eqdilog1}, we
obtain an expression of the form $-\frac{1}{2\lambda}[[\cdots]]$, where
the double square brackets enclose
\begin{multline*}
	-\ln\left|c^2-T^2\right| (t-s) \left(\ln|t-s| -1 \right) - (t-s) \\
	+ \left((t - s)  - \frac{c+T}{2\lambda} \right) \ln \left\vert 1- \frac{2\lambda (t-s)}{c+T} \right\vert  \\
	- (t -s ) \LL \left(\frac{c+T}{2\lambda (t-s)} \right) - \frac{(t-s)}{2} \left(\ln \left|\frac{c+T}{2\lambda (t-s)}\right|\right)^2 \\
	+ \frac{\pi^2}{12}(t-s) +  \frac{\pi^2}{4} (t-s) \sgn\left(\frac{c+T}{2\lambda (t-s)}\right) .
\end{multline*}
Evaluating this result at the boundaries $(s,t)=(0,0)$ and $(s,t)=(1,1)$
which correspond to the $z_{11}$ and $z_{22}$ vertices gives zero. The
$(s,t)=(1,0)$ and $(s,t)=(0,1)$ boundaries which map to the $z_{12}$ and
$z_{21}$ vertices give a nonzero result given by
\begin{multline*}
	- \frac{1}{2\lambda} \Biggl(\frac{\ln|1+\frac{2\lambda}{c+T}|}{\ln|1-\frac{2\lambda}{c+T}|} + \frac{(c+T)}{2\lambda}\ln\biggl| 1-\left(\frac{2\lambda}{c+T}\right)^2\biggr| \\
	+ \LL\left(\frac{c+T}{2\lambda}\right) - \LL\left(-\frac{c+T}{2\lambda}\right) - \frac{\pi^2}{2} \sgn\left(\frac{c+T}{2\lambda}\right) \Biggr).
\end{multline*}
Expanding this for small $c+T$ yields
\begin{multline*}
	-\frac{1}{2\lambda}\left(2\frac{c+T}{\lambda} - \frac{c+T}{\lambda} \ln\left|\frac{c+T}{2\lambda}\right| - \frac{\pi^2}{2} \sgn\left(\frac{c+T}{2\lambda}\right)\right) \\
	+ \O(c+T)^2,
\end{multline*}
where we recall that the summation over $c=\pm R$ still needs to be
performed. Integration over $\phi$ and $T$ is trivial and using the
definition in Eq.~\eqref{eqdefmu}, the result is 
\begin{equation*}
\tilde{I}^{00}=\frac{\pi}{\lambda^2}\left(4 + 2 \ln|2 \lambda/\mu| - \frac{\pi^2 \, \lambda}{\mu}\right) + \O(\mu^0).
\end{equation*}
The same calculation was also checked by hand using the momentum space
representation of the Hadamard two-point function
Eq.~\eqref{eqgrav2ptmomentum}, with the smearing and $s,t$ integrals
also converted to momentum space. The result agreed with the above,
giving us confidence that it is correct. This result is also reproduced
by the computer calculation, giving us confidence that the latter is
correct as well.

The same procedure can also be used to calculate terms with one or two
derivatives on the smearing function. However, these calculations are
slightly more involved since now also terms proportional to $c$ and
$c^2$ appear and integration by parts is needed. The results of the
calculation for these integrals have been checked by hand and are quoted
here without any intermediate steps. These results also agree with the
computer output. For one derivative on the smearing function ($|K|=1$)
we have $\tilde{I}^{00}_{u} = \tilde{I}^{00}_{\u} = 0$.  For two
derivatives on the smearing function ($|K|=2$), the complete set of
smeared segment integrals is
\begin{align*}
&\tilde{I}^{00}_{uu} =  &\frac{\pi}{\lambda^2}\left(-\frac{2}{\mu^2} + \pi^2 \, \lambda \, g(0) \right) + \O(\mu^0),\\
&\tilde{I}^{00}_{u\u} =  &\frac{\pi}{\lambda^2}\left(-\frac{8}{\mu^2} -\frac{16 \ln|2 \lambda/\mu|}{\mu^2} + 4 \pi^2 \, \lambda \, g(0) \right) + \O(\mu^0),\\
&\tilde{I}^{00}_{\u\u} =  &\frac{\pi}{\lambda^2}\left(\frac{10}{\mu^2} -\frac{12 \ln|2 \lambda/\mu|}{\mu^2} + \pi^2 \, \lambda \, g(0) \right) + \O(\mu^0),\\
&\tilde{I}^{00}_{\delta_\perp} =  &\frac{\pi}{\lambda^2}\left(-\frac{4}{\mu^2} -\frac{4 \ln|2 \lambda/\mu|}{\mu^2} \right)+ \O(\mu^0).
\end{align*}

\bibliographystyle{apsrev4-1}
\bibliography{paper-delay2}

\end{document}